\documentclass[12pt]{article}

\usepackage[utf8]{inputenc}
\usepackage{float,bm,enumerate,relsize,empheq,graphicx,amssymb,amsmath,amsthm,amsfonts,fancyhdr,cite}
\numberwithin{equation}{section}
\usepackage[lmargin=2.5cm,rmargin=2.5cm,tmargin=2.5cm,bmargin=2.5cm]{geometry} 
\usepackage[usenames,dvipsnames]{color}
\begin{document}
\pagestyle{fancy}

\lhead{SUSY QM and Painlev\'e equations}
\rhead{D. Bermudez, D.J. Fern\'andez C.}

\title{Supersymmetric quantum mechanics\\ and Painlev\'e equations}
\author{David Bermudez\footnote{{\it email:} dbermudez@fis.cinvestav.mx},\, 
David J. Fern\'andez C.\footnote{{\it email:} david@fis.cinvestav.mx} \\
[8pt]
{\sl Departamento de F\'{\i}sica, Cinvestav}\\{\sl A.P. 14-740, 07000 M\'exico D.F., Mexico}}

\date{}

\maketitle

\begin{abstract}
In these lecture notes we shall study first the supersymmetric quantum mechanics (SUSY QM), specially when applied to the harmonic and radial oscillators. In addition, we will define the polynomial Heisenberg algebras (PHA), and we will study the general systems ruled by them: for zero and first order we obtain the harmonic and radial oscillators, respectively; for second and third order PHA the potential is determined by solutions to Painlev\'e IV (PIV) and Painlev\'e V (PV) equations. Taking advantage of this connection, later on we will find solutions to PIV and PV equations expressed in terms of confluent hypergeometric functions. Furthermore, we will classify them into several {\it solution hierarchies}, according to the specific special functions they are connected with.

\end{abstract}

\section{Introduction}
Through the years, there has been different connections between quantum mechanics and non-linear differential equations. The simplest case was studied by Dirac \cite{Dir30} between Schr\"odinger and Riccati equations, a second-order linear and first-order non-linear differential equations, respectively. Further examples are the SUSY partners of the free particle potential, which lead to solutions of the Korteweg-de Vries (KdV) equation \cite{Mat92}.

In these lecture notes we will study in detail the relation between supersymmetric quantum mechanics (SUSY QM), polynomial Heisenberg algebras (PHA), and Painlev\'e equations. It has been known for years that specific PHA were connected with solutions of some Painlev\'e equations and that the first-order SUSY partner potentials of the harmonic and radial oscillator were ruled by these algebras. Now, several questions arise: can higher-order SUSY partners lead to additional solutions? And if so, which are the corresponding conditions imposed on the quantum systems? What kind of solutions do they lead to? In this work we intent to answer all these questions.

We will see that the second and third-order PHA are related with Painlev\'e IV (PIV) and Painlev\'e V (PV) equations respectively. Not only that, but we will use higher-order SUSY QM to obtain systems described by these two kinds of algebras departing from the harmonic and the radial oscillators, which will allow us to find new solutions of PIV and PV equations. After that, we will classify these solutions into the so called {\it solution hierarchies}, according to the specific special functions they are related with.

As far as we know, the first people who realized the connection between SUSY QM, second and third-order PHA, and 
Painlev\'e equations were Veselov and Shabat \cite{VS93}, Adler \cite{Adl93}, and Duvov et al. \cite{DEK94}. This connection has been explored more thoroughly by Fern\'andez, Negro, Nieto, and Mateo \cite{ACIN00,FNN04,CFNN04,MN08} and by Bermudez and Fern\'andez \cite{Ber10,BF11a,BF11b,Ber12,BF12,BF13a}.

These lecture notes are organized as follows. In section 2 we will review briefly the SUSY QM, while in section 3 we will introduce the six Painlev\'e equations. In section 4 we will define the PHA and in section 5 we will describe the general systems characterized by them for the four lowest orders. Then, in section 6 we will introduce our method to obtain solutions to PIV equation and in section 7 we will do the same for PV: in each section we will present a reduction theorem which will guarantee the corresponding connection, as well as a classification of the solutions into hierarchies. Finally, we will present our conclusions in section 8.

\section{Supersymmetric quantum mechanics}
\label{capsusyqm}

The factorization method, intertwining technique, and SUSY QM are closely related and their names will be used indistinctly in this work to characterize a specific method, through which it is possible to obtain new exactly-solvable quantum potentials departing from known ones.

\subsection{First-order SUSY QM}
\label{secsusy1}
Let $H_0$ and $H_1$ be two Schr\"odinger Hamiltonians
\begin{equation}
H_j = -\frac{1}{2}\frac{\text{d}^2}{\text{d}x^2} + V_j (x) \text{, \ \ \ \ \ } j=0,1.
\end{equation}
For simplicity, we are taking {\it natural units} $\hbar=m=1$. Next, let us suppose the existence of a first-order differential operator $A_1^{+}$ that {\it intertwines} the two Hamiltonians in the way
\begin{equation}
H_1 A_1^{+}=A_1^{+}H_0, \qquad A_1^{+}=\frac{1}{2^{1/2}}\left[-\frac{\text{d}}{\text{d}x} + \alpha_1 (x)\right],\label{entre}
\end{equation}
where the \emph{superpotential} $\alpha_1 (x)$ is still to be determined. We must remind that these equations involve operators, which implies that in order to interchange the differential operator $\text{d}^k/\text{d}x^k$ with any multiplicative operator function $f(x)$ we must use the following relations
\begin{subequations}
\begin{align}
\frac{\text{d}}{\text{d}x} f &=f\frac{\text{d}}{\text{d}x} + f',\\
\frac{\text{d}^2}{\text{d}x^2} f &=f\frac{\text{d}^2}{\text{d}x^2} +2f'\frac{\text{d}}{\text{d}x} +f''.\\
		&\ \ \ \ \ \vdots\nonumber
\end{align}
\end{subequations}

Then, for both sides of equation \eqref{entre} it is straightforward to show that
\begin{subequations}
\begin{align}
2^{1/2}H_1 A_1^{+} &=\frac{1}{2}\frac{\text{d}^3}{\text{d}x^3}-\frac{\alpha_1}{2}\frac{\text{d}^2}{\text{d}x^2}-(V_1 + \alpha'_1)\frac{\text{d}}{\text{d}x} + \alpha_1 V_1 - \frac{\alpha''_1}{2},\\
2^{1/2}A_1^{+}H_0 &= \frac{1}{2}\frac{\text{d}^3}{\text{d}x^3}-\frac{\alpha_1}{2}\frac{\text{d}^2}{\text{d}x^2}-V_0 \frac{\text{d}}{\text{d}x} +\alpha_1 V_0 -V'_0.
\end{align}\label{AH}
\end{subequations}
\hspace{-1.5mm}Matching the powers of $\text{d}/\text{d}x$ in equations \eqref{AH} and solving the coefficients, we get
\vspace{-2mm}
\begin{subequations}
\begin{align}
V_1(x) &= V_0(x) - \alpha'_1(x,\epsilon),\label{Valfa}\\
\alpha'_1(x,\epsilon)+\alpha^{2}_1(x,\epsilon) &= 2[V_0(x)-\epsilon]. \label{alfa}
\end{align}\label{2alfas}
\end{subequations}
\vspace{-2mm}
\hspace{-1.5mm}If we define $u^{(0)}(x)$ such that $\alpha_1(x,\epsilon)=u^{(0)'}/u^{(0)}$, then equations \eqref{2alfas} are mapped into
\begin{subequations}
\begin{align}
V_1 &=  V_0- \left[\frac{u^{(0)'}}{u^{(0)}}\right]',\label{Vu}\\
-\frac{1}{2}u^{(0)''} + V_0 u^{(0)} &= \epsilon u^{(0)},\label{ucero}
\end{align}\end{subequations}
which means that $u^{(0)}$ is a solution of the initial stationary Schr\"odinger equation associated with $\epsilon$, although it might not fulfill any boundary condition.

Starting from equations \eqref{2alfas} we obtain that $H_0$ and $H_1$ can be factorized as
\begin{equation}
H_0 = A_1^{-} A^{+}_1+\epsilon,\qquad
H_1 = A^{+}_1 A_1^{-}+\epsilon, \label{ffH}
\end{equation}
where
\begin{equation}
A_1^{-}\equiv(A^{+}_1)^{\dagger}=\frac{1}{2^{1/2}}\left[\frac{\text{d}}{\text{d}x} + \alpha_1(x,\epsilon)\right].\label{amam}
\end{equation}

\begin{figure}[H]
\centering
\includegraphics[scale=0.29]{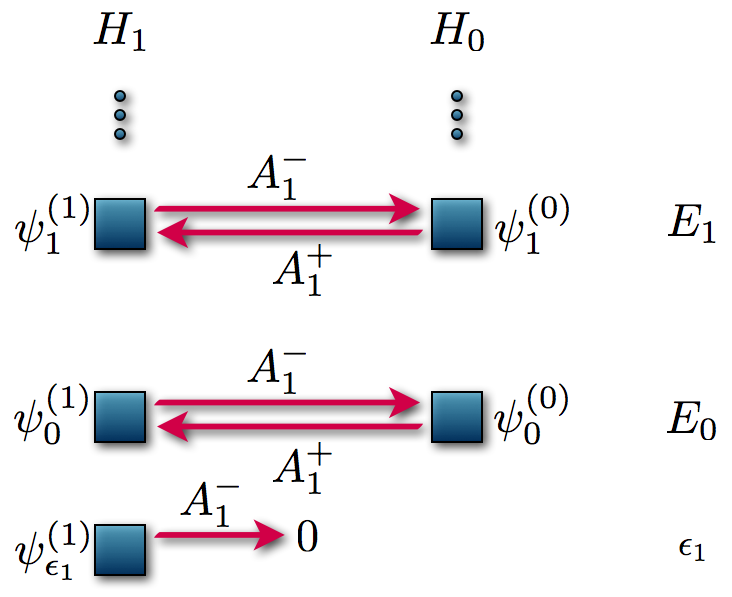}
\caption{\small{Diagram of the first-order SUSY transformation. The final Hamiltonian $H_1$ has the same spectrum of the initial one $H_0$ plus a new level at the factorization energy $\epsilon_1$.}}
\label{fig.susyqm1er}
\end{figure}

\begin{figure}[H]
\centering
\includegraphics[scale=0.48]{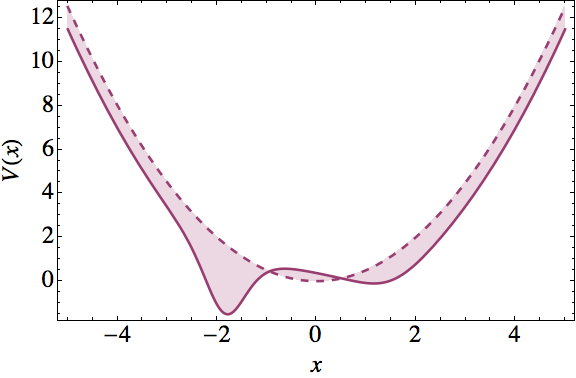}
\caption{\small{First-order SUSY partner $V_1(x)$ (solid line) of the harmonic oscillator potential $V_0(x)$ (dashed 
line), generated using $u^{(0)}(x)$ with $\epsilon=0,\ \nu=0.9$. We remark the difference between the two potentials.}}\label{figsusy1}
\end{figure}

Let us assume that $V_0(x)$ is a solvable potential with normalized eigenfunctions $\psi^{(0)}_n (x)$ and eigenvalues 
$E_n, \ n=0,1,2,\dots$. Besides, we know a non-singular solution $\alpha_1(x,\epsilon)$ [$u^{(0)}(x)$ without zeroes] to the Riccati equation \eqref{alfa} [Schr\"odinger \eqref{ucero}] for $\epsilon=\epsilon_1 \leq E_0$, where $E_0$ is the ground state energy for $H_0$. Then, the potential $V_1(x)$ given in equation \eqref{Valfa} [\eqref{Vu}] is determined, and its normalized eigenfunctions are
\begin{subequations}
\begin{align}
\psi^{(1)}_{\epsilon_1}(x) &\propto \exp\left(-\int^{x}_0 \alpha_1(y,\epsilon_1)\text{d}y\right)=\frac{1}{u^{(0)}_1(x)},\label{psimathcal}\\
\psi^{(1)}_{n}(x) &= \frac{A^{+}_{1} \psi_n^{(0)}(x)}{(E_n-\epsilon_1)^{1/2}},
\end{align}\label{psis}
\end{subequations}
\hspace{-1.5mm}while its eigenvalues are such that Sp$(H_1)=\{\epsilon_1,E_n,n=0,1,\dots\}$. An scheme of the way the first-order SUSY transformation works, and the resulting spectrum, is shown in figure \ref{fig.susyqm1er}. An example of the generated potentials can be seen in figure \ref{figsusy1}.

\medskip

\subsection{Higher-order SUSY QM}
Let us apply this technique iteratively, taking now the resulting $V_1(x)$ as a solvable potential which is used to generate a new one $V_2(x)$ through another intertwining operator $A^{+}_2$ and a different factorization energy $\epsilon_2$, with the restriction $\epsilon_2<\epsilon_1\leq 1/2$ once again taken to avoid singularities in the new potential and its eigenfunctions. The corresponding intertwining relation reads
\begin{equation}
H_2A^{+}_2=A^{+}_2H_1,
\end{equation}
which leads to equations similar to \eqref{2alfas} for $V_2$ and $\alpha_2$:
\vspace{-2mm}
\begin{subequations}
\begin{align}
V_2(x) &= V_1(x) - \alpha'_2(x,\epsilon_2),\\
\alpha'_2(x,\epsilon_2)+\alpha^{2}_2(x,\epsilon_2) &= 2[V_1(x)-\epsilon_2]. \label{alfa2}
\end{align}
\end{subequations}
\hspace{-1mm}In terms of $u^{(1)}_2(x)$ such that $\alpha_2 (x,\epsilon_2)=u_2^{(1)}{}'(x)/u^{(1)}_2(x)$ we have
\begin{equation}
V_2  =  V_1 -\left[\frac{u_2^{(1)}{}'}{u^{(1)}_2}\right]',\qquad 
-\frac{1}{2}u_2^{(1)}{}''+V_1 u^{(1)}_2 = \epsilon_2 u^{(1)}_2.
\end{equation}

An important result that will be proven next is that the solution of the Riccati equation \eqref{alfa2} for $\alpha_2$ can be algebraically determined using the solutions of the initial Riccati equation \eqref{alfa} for the factorization energies $\epsilon_1$ and $\epsilon_2$ \cite{FHM98,Ros98a,Ros98b,FH99,MNR00}. To do that, first let us take the two solutions of the initial Riccati equation 
\begin{equation}
\alpha'_1(x,\epsilon_j)+\alpha^{2}_1(x,\epsilon_j) = 2[V_0(x)-\epsilon_j], \qquad j=1,2. \label{alfas}
\end{equation}
Therefore, for the Schr\"odinger equation we have $H_0u^{(0)}_j(x)=\epsilon_j u^{(0)}_j(x)$, where
\begin{equation}
u^{(0)}_j(x) \propto \exp\left(\int^{x}_0 \alpha_1 (y,\epsilon_j)\text{d}y\right).
\end{equation}

Let us recall that $u^{(0)}_1(x)$ is used to implement the first transformation so that the eigenfunction of $H_1$ associated with $\epsilon_1$ is given by equation~\eqref{psimathcal} and the one with $\epsilon_2$ is
\begin{equation}
u^{(1)}_2 \propto A^{+}_1 u^{(0)}_2 \propto - u_2^{(0)}{}'+\alpha_1 (x,\epsilon_1)u^{(0)}_2 \propto \frac{W\left(  u^{(0)}_1,u^{(0)}_2 \right) }{u^{(0)}_1}.
\end{equation}
Taking into account that $u_2^{(0)}{}' = \alpha_1 (x,\epsilon_2)u^{(0)}_2$ we have
\begin{equation}
u^{(1)}_2 \propto [\alpha_1(x,\epsilon_1)-\alpha_1(x,\epsilon_2)]u^{(0)}_2.\label{u2u1}
\end{equation}

To implement the second transformation we express now $u^{(1)}_2$ in terms of the corresponding superpotential
\begin{equation}
u^{(1)}_2(x) \propto \exp\left(\int^{x}_0 \alpha_2 (y,\epsilon_2)\text{d}y\right). \label{u2alfa}
\end{equation}
Substituting \eqref{u2alfa} in equation~\eqref{u2u1} we obtain
\begin{equation}
\exp\left(\int^{x}_0 \alpha_2 (y,\epsilon_2)\text{d}y\right) \propto [\alpha_1(x,\epsilon_1)-\alpha_1(x,\epsilon_2)]u^{(0)}_2(x).
\end{equation}
Taking the logarithm on both sides and deriving with respect to $x$:
\begin{equation}
\alpha_2 (x,\epsilon_2) = \alpha_1(x,\epsilon_2)+\frac{\alpha'_1 (x,\epsilon_1)-\alpha'_1(x,\epsilon_2)}{\alpha_1 (x,\epsilon_1)-\alpha_1(x,\epsilon_2)}.
\end{equation}
Using the initial Riccati equations \eqref{alfas} to eliminate the derivatives of $\alpha_1$ we obtain
\begin{equation} \alpha_2(x,\epsilon_2)=-\alpha_1(x,\epsilon_1)-\frac{2(\epsilon_1-\epsilon_2)}{\alpha_1(x,\epsilon_1)-\alpha_1(x,\epsilon_2)}.\label{alfa2b}
\end{equation}

This formula expresses the solution of the Riccati equation (\ref{alfa2}) with $V_1(x)=V_0(x)-\alpha'_1(x,\epsilon_1)$ as a {\it finite difference formula} that involves two solutions, $\alpha_1(x,\epsilon_1)$ and $\alpha_1(x,\epsilon_2)$, of the Riccati equation \eqref{alfa} for the factorization energies $\epsilon_1,\epsilon_2$ \cite{FHM98}.

\begin{figure}\centering
\includegraphics[scale=0.34]{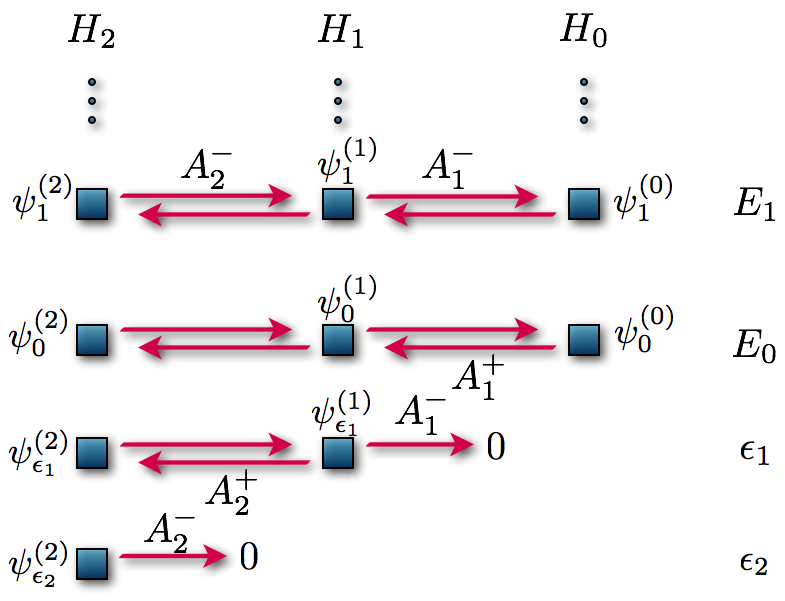}
\caption{\small{Iteration of two first-order SUSY transformations applied to $H_0$ and $H_1$ using as transformation functions two non-physical eigenfunctions of $H_0$ with factorization energies $\epsilon_2 < \epsilon_1 < E_0$.}}\label{fig.susyqm2do}
\end{figure}

On the other hand, the potential $V_2(x)$ is expressed as
\begin{equation}
V_2(x)=V_1(x)-\alpha'_2 (x,\epsilon_2)=V_0(x) +\left[\frac{2(\epsilon_1-\epsilon_2)}{\alpha_1(x,\epsilon_1)-\alpha_1(x,\epsilon_2)}\right]',
\end{equation}
the eigenfunctions associated with $H_2$ are given by
\begin{subequations}
\begin{align}
\psi^{(2)}_{\epsilon_2}(x) &\propto \exp\left(-\int^{x}_0 \alpha_2(y,\epsilon_2)\text{d}y\right)=\frac{1}{u^{(1)}_2(x)}
\propto \frac{u^{(0)}_1}{W\left(  u^{(0)}_1,u^{(0)}_2 \right) }, \label{psi22a}\\
\psi^{(2)}_{\epsilon_1}(x) &= \frac{A^{+}_{2} \psi^{(1)}_{\epsilon_1}(x)}{(\epsilon_1-\epsilon_2)^{1/2}},\\
\psi^{(2)}_{n}(x) &= \frac{A^{+}_{2} \psi^{(1)}_{n}(x)}{(E_n-\epsilon_2)^{1/2}}= \frac{A^{+}_2A^{+}_1\psi^{(0)}_n(x)}{[(E_n-\epsilon_1)(E_n-\epsilon_2)]^{1/2}},
\end{align}
\end{subequations}
and the corresponding eigenvalues are such that Sp$(H_2)=\{\epsilon_2,\epsilon_1,E_n,n=0,1,\dots\}$. A scheme representing this transformation is shown in figure \ref{fig.susyqm2do}.

This iterative process can be continued to higher orders. Thus, let us assume that we know $k$ solutions $\{ \alpha_1(x,\epsilon_j), j=1,\dots,k\}$ of the initial Riccati equation, where $\epsilon_{j+1}<\epsilon_{j}$. Therefore, we obtain a new solvable Hamiltonian $H_k$, whose potential reads
\begin{equation}
V_k(x)=V_{k-1}(x)-\alpha'_k(x,\epsilon_k)=V_0(x) - \sum_{j=1}^{k}\alpha'_j(x,\epsilon_j),\label{Vk}
\end{equation}
where $\alpha_j(x,\epsilon_j)$ is given by a recursive finite difference formula which is obtained as a generalization of equation \eqref{alfa2b}, i.e.,
\begin{equation}
\alpha_{j+1}(x,\epsilon_{j+1})=-\alpha_j(x,\epsilon_j)-\frac{2(\epsilon_j-\epsilon_{j+1})}{\alpha_j(x,\epsilon_j)-\alpha_j(x,\epsilon_{j+1})}, \ \ \ j=1,\dots,k-1. \label{alfai}
\end{equation}

The eigenfunctions of $H_k$ are given by
\begin{subequations}
\begin{align}
\psi^{(k)}_{\epsilon_k}(x) &\propto \exp\left(-\int^{x}_0 \alpha_k(y,\epsilon_k)\text{d}y\right),\\
\psi^{(k)}_{\epsilon_{k-1}}(x) &= \frac{A^{+}_{k}\psi^{(k-1)}_{\epsilon_{k-1}}(x)} {(\epsilon_{k-1}-\epsilon_k)^{1/2}},\label{psie}\\
				&\hspace{2mm} \vdots \nonumber\\
\psi^{(k)}_{\epsilon_1}(x) &= \frac{A^{+}_{k}\dots A^{+}_2 \psi^{(1)}_{\epsilon_1}(x)} {[(\epsilon_1-\epsilon_2)\dots (\epsilon_1-\epsilon_k)]^{1/2}},\\
\psi^{(k)}_{n}(x) &=\frac{A^{+}_k\dots A^{+}_1\psi^{(0)}_n(x)} {[(E_n-\epsilon_1)\dots(E_n-\epsilon_k)]^{1/2}}.\label{psin}
\end{align}\label{psisk}
\end{subequations}
\hspace{-1.5mm}The corresponding eigenvalues belong to the set Sp$(H_k)=\{ \epsilon_j, E_n, j=k,\dots,1,$ $n=0,1,\dots \}$.

In order to complete our scheme, let us recall how the SUSY partners $H_{j-1}$, $H_j$ are intertwined
\begin{equation}
H_j A^{+}_j=A_j^{+}H_{j-1}, \qquad j=1,\dots ,k. \label{HAAH}
\end{equation}
Then, starting from $H_0$ we have generated a chain of factorized Hamiltonians in the way
\begin{equation}
H_j=A^{+}_jA_j^{-}+\epsilon_j,\qquad H_{j-1}=A^{-}_{j}A^{+}_{j}+\epsilon_{j}, \qquad j=1,\dots,k,
\label{HAA}
\end{equation}
\hspace{-1.5mm}where the final potential $V_k(x)$ can be determined using equations \eqref{Vk} and \eqref{alfai}.

In addition, since we are departing from $k$ solutions of the initial Riccati equation, $\{\alpha_1(x,\epsilon_i);\ i=1,\dots,k\}$, then we also obtain $k$ non-equivalent factorizations of the Hamiltonian $H_0$,
\begin{equation}
H_0 = \frac{1}{2}\left[\frac{\text{d}}{\text{d}x} + \alpha_1(x,\epsilon_i)\right]\left[-\frac{\text{d}}{\text{d}x}+\alpha_1(x,\epsilon_i)\right]+\epsilon_i, \qquad i=1,\dots,k.
\end{equation}

We must note now that there exists a $k$th-order differential operator, $B_{k}^{+} \equiv A^{+}_k\dots A^{+}_1$, that intertwines the initial and final Hamiltonians $H_0$ and $H_k$ as follows
\begin{equation}
H_k B^{+}_k = B^{+}_k H_0.
\label{Bdag}
\end{equation}

From equations \eqref{psisk} and the adjoint of \eqref{Bdag} we arrive to
\begin{subequations}
\begin{align}
B^{+}_k\psi^{(0)}_n &= [(E_n-\epsilon_1)\dots(E_n-\epsilon_k)]^{1/2}\psi^{(k)}_n,\label{Bdag2}\\
B_k^{-}\psi^{(k)}_n &= [(E_n-\epsilon_1)\dots(E_n-\epsilon_k)]^{1/2}\psi^{(0)}_n.\label{Bdag3}
\end{align}
\end{subequations}

These equations immediately lead to the higher-order SUSY QM \cite{FHM98,Ros98a,Ros98b,FH99,MNR00,AIS93,AICD95,BS97,FGN98,BGBM99,mr04,ff05,fe10,ai12}. In this treatment, the standard SUSY algebra with two generators $\{Q_1,Q_2\}$ \cite{Wit81},
\begin{equation}
[Q_i,H_{ss}]=0, \ \ \ \{Q_i,Q_j\}=\delta_{ij}H_{ss}, \ \ \ i,j=1,2,
\label{Qis}
\end{equation}
can be realized from $B_k^{-}$ and $B^{+}_k$ through the definitions
\begin{equation}
Q^{-}=\left(\begin{array}{cc}
  0 &  0 \\
  B_k^{-} &  0 
  \end{array}\right),\quad 
Q^{+}=\left(\begin{array}{cc}
  0 &  B^{+}_k \\
  0 &  0 
  \end{array}\right),\quad
H_{ss}\equiv\{Q^{-},Q^{+}\}=\left(\begin{array}{cc}
  B^{+}_kB_k^{-} &  0 \\
  0 &  B_k^{-}B^{+}_k 
  \end{array}\right),
  \label{Qis2}
\end{equation}
where $Q_1\equiv (Q^{+}+Q^{-})/2^{1/2}$ and $Q_2\equiv i(Q^{-}-Q^{+})/2^{1/2}$. Given that
\begin{subequations}
\begin{align}
B^{+}_kB_k^{-}&=A^{+}_k\dots A^{+}_1 A_1^{-}\dots A_k^{-}=(H_k-\epsilon_1)\dots (H_k-\epsilon_k), \label{Bdag4}\\
B_k^{-}B^{+}_k			&=A_1^{-}\dots A_k^{-} A^{+}_k\dots A^{+}_1=(H_0-\epsilon_1)\dots (H_0-\epsilon_k),\label{Bdag5}
\end{align}
\end{subequations}
it turns out that the SUSY generator ($H_{ss}$) is a $k$th-order polynomial of the Hamiltonian $H^p_s$ that involves the two intertwined Hamiltonians $H_0$ and $H_k$,
\begin{equation}
H_{ss}=(H^p_s -\epsilon_1)\dots(H^p_s -\epsilon_k),
\label{Hss}
\end{equation}

where
\begin{equation}
H^p_s=\left(\begin{array}{cc}
  H_k &  0 \\
  0 &  H_0 
  \end{array}\right).\\
  \label{Hps}
\end{equation}

\subsubsection*{Example 1. The harmonic oscillator}
Let us consider first the harmonic oscillator potential $V_0(x)=x^2/2$. Its eigenfunctions and eigenvalues can be found through the factorization method, which is based on the fact that
\begin{equation}
H_0 = a^- a^+ - \frac12 = a^+ a^- + \frac12,
\end{equation}
where $a^-$ and $a^+$ are the annihilation and creation operators given by
\begin{equation}
a^- = \frac{1}{2^{1/2}}\left(\frac{\text{d}}{\text{d}x} + x \right),  \qquad 
a^+ = \frac{1}{2^{1/2}}\left(-\frac{\text{d}}{\text{d}x} + x \right). 
\end{equation}
It is important to define also the number operator
\begin{equation}
N \equiv a^+ a^-.
\end{equation}
The operator set $\{N,a^-, a^+\}$ satisfies the following commutation relations
\begin{equation}
[N,a^\pm] = \pm a^\pm, \qquad [a^-, a^+] = 1,
\end{equation}
where $1$ denotes here the identity operator. In Lie algebraic terminology, it is said that $\{N,a^-, a^+\}$ generate de Heisenberg-Weyl algebra. 

From these relations, it is straightforward to derive the normalized eigenfunctions $\psi_n(x)$ and eigenvalues $E_n$ of $H_0$ departing from the normalized ground state eigenfunction $\psi_0(x)$,
\begin{equation}
\psi_n(x) = (n!)^{-1/2} (a^+)^n \psi_0(x), \quad E_n = n + \frac12, \quad n=0,1,2,\dots
\end{equation}
Note that $\psi_0(x)$ is annihilated by $a^-$; its explicit expression is given by
\begin{equation}
\psi_0(x) = \pi^{-1/4} \exp\left(-\frac{x^2}{2}\right).
\end{equation}
For completeness, let us write down the action of $a^-$ and $a^+$ onto $\psi_n(x)$:
\begin{equation}
a^-\psi_n(x) = \sqrt{n}\,\psi_{n-1}(x), \qquad a^+\psi_n(x) = \sqrt{n+1}\,\psi_{n+1}(x).
\end{equation}

We can proceed now to implement the SUSY 	treatment for the harmonic oscillator. In order to apply the first-order SUSY transformation, we just need to supply either a non-singular solution of the Riccati equation \eqref{alfa} or a nodeless one of the Schr\"odinger equation \eqref{ucero}. The general solution of the stationary Schr\"odinger equation for the harmonic oscillator potential with $\epsilon\in\mathbb{R}$ is given by
\begin{equation}
u(x)=\exp(-x^2/2)\left[ {}_1F_1\left(\frac{1-2\epsilon}{4},\frac{1}{2};x^2\right)+2\nu\frac{\Gamma\left(\frac{3-2\epsilon}{4}\right)}{\Gamma\left(\frac{1-2\epsilon}{4}\right)}\,x\, {}_1F_1\left(\frac{3-2\epsilon}{4},\frac{3}{2};x^2\right)\right],
\label{hyper}
\end{equation}
where
\begin{equation}
{}_1F_1(a,b;y)= \frac{\Gamma(b)}{\Gamma(a)}\sum_{n=0}^{\infty}\frac{\Gamma(a+n)}{\Gamma(b+n)}\frac{y^n}{n!},
\end{equation}
is the confluent hypergeometric function, $\Gamma(x)$ is the Gamma function and $\nu\in\mathbb{R}$. Thus, the first-order SUSY partner potential $V_1(x)$ of the harmonic oscillator becomes
\begin{equation}
V_1(x)=\frac{x^2}{2}-[\ln u(x)]''.
\end{equation}
Note that, for $\epsilon < 1/2$ this solution will not have zeroes for $\vert\nu\vert < 1$ while it will have only one for $\vert\nu\vert > 1$; this means that the nodeless solution $u(x)$ is chosen so that $\epsilon < 1/2, \ \vert\nu\vert <1$.

Let us perform now a non-singular $k$th-order SUSY transformation which creates precisely $k$ new
levels, additional to $E_n = n + 1/2, \ n=0,1,2,\dots$ of $H_0$, in the way
\begin{equation}
\text{Sp}(H_k)  = \left\{ \epsilon_k, \dots,\epsilon_1, \frac12,\frac32,\dots \right\} , \label{spect}
\end{equation}
where $\epsilon_k < \cdots < \epsilon_1 < 1/2$. In order that the Wronskian $W(u_1,\dots,u_k)$ would be nodeless, the parameters $\nu_j$ have to be taken as $\vert\nu_j\vert < 1$ for $j$ odd and $\vert\nu_j\vert > 1$ for $j$ even, $j=1,\dots,k$. The corresponding potential turns out to be
\begin{equation}
V_k(x) = \frac{x^2}2 - \{\ln[W(u_1,\dots,u_k)]\}''.
\end{equation}

It is important to note that there is a pair of natural ladder operators $L_k^\pm$ for $H_k$, i.e., $L_k^{\pm} = B_k^{+} a^{\pm} B_k^{-},$ which are differential operators of $(2k+1)$-th order such that $[H_k,L_k^\pm]= \pm L_k^\pm$. From equations \eqref{Bdag}, \eqref{Bdag4}, and \eqref{Bdag5}, it is straightforward to show the following relation in terms of the {\it extremal energies} $\mathcal{E}_j$
\begin{equation}
N(H_k) = L_k^+ L_k^- =\prod_{j=1}^{2k+1}(H_k-\mathcal{E}_j)
= \left(H_k - \frac12\right) \prod_{j=1}^k \left(H_k - \epsilon_j \right) \left(H_k - \epsilon_j - 1\right). \label{annum}
\end{equation}

\subsubsection*{Example 2. The radial oscillator}\label{SUSYRO}
Now we will apply the $k$-SUSY QM to the radial oscillator Hamiltonian, which is given by
\begin{equation}
H_\ell=-\frac{1}{2}\frac{\text{d}^2}{\text{d}x^2}+\frac{x^2}{8}+\frac{\ell(\ell+1)}{2x^2}, \qquad \ell\geq 0, \quad x\geq 0,
\end{equation}
where we have added the subscript $\ell$ to denote the dependence of the Hamiltonian in the angular momentum index.

To perform the higher-order SUSY transformations onto this potential we will explore once again the factorization method \cite{CFNN04,FND96,CR08}. The radial oscillator Hamiltonian can be factorized directly in four different ways. The first two are
\begin{equation}
H_\ell=a_\ell^- a_\ell^+ +\frac{\ell}{2}-\frac{1}{4} = a_{\ell+1}^+a_{\ell+1}^-+\frac{\ell}{2}+\frac{3}{4},
\end{equation}
with
\begin{equation}
a_\ell^\pm \equiv \frac{1}{2^{1/2}}\left(\mp\frac{\text{d}}{\text{d}x}-\frac{\ell}{x}+\frac{x}{2}\right).
\end{equation}

The commutator of these operators is
\begin{equation}
[a_\ell^-,a_\ell^+]=\frac{\ell}{x^2}+\frac{1}{2},
\end{equation}
from which we can see that $a_\ell^\pm$ are not ladder operators but rather {\it shift operators}, i.e., they change the angular momentum of $H_\ell$ and intertwine it with $H_{\ell-1}$, creating a hierarchy of Hamiltonians with different $\ell$
\begin{equation}
H_\ell a_\ell^- = a_\ell^-\left(H_{\ell-1}-\frac{1}{2}\right),\qquad H_{\ell-1} a_\ell^+ = a_\ell^+\left(H_{\ell}+\frac{1}{2}\right).\label{interRO}
\end{equation}

Now, let $\psi_{n\ell}(x)$ be an eigenfunction of $H_\ell$ with eigenvalue $E_{n\ell}$, i.e., $H_\ell \psi_{n\ell} = E_{n\ell}\psi_{n\ell}$. Then, from equations~\eqref{interRO} we obtain that
\begin{equation}
H_{\ell+1} (a_{\ell+1}^- \psi_{n\ell}) = \left(E_{n\ell}-\frac{1}{2}\right)(a_{\ell+1}^-\psi_{n\ell}),\qquad
H_{\ell-1} (a_{\ell}^+ \psi_{n\ell}) = \left(E_{n\ell}+\frac{1}{2}\right)(a_{\ell}^+\psi_{n\ell}).
\end{equation}

\begin{figure}
\centering
\includegraphics[scale=0.34]{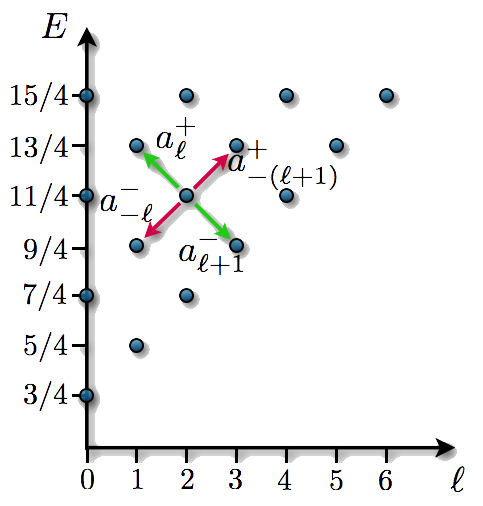}
\includegraphics[scale=0.34]{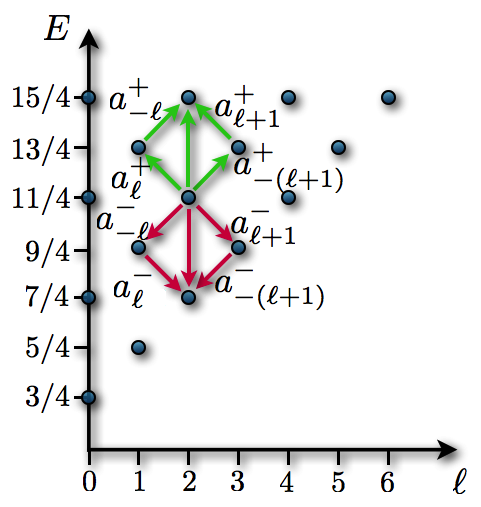}
\caption{\small{Diagram of the action of the first-order shift operators $a_\ell^\pm$  and $a_{-\ell}^\pm$ (left). On the horizontal axis we have the angular momentum index $\ell$ and on the vertical one the energy values. The joint action of two appropriate shift operators leads to the second-order ladder operators $b_\ell^\pm$ (right).}}\label{diaRO}
\end{figure}

\begin{figure}
\centering
\includegraphics[scale=0.43]{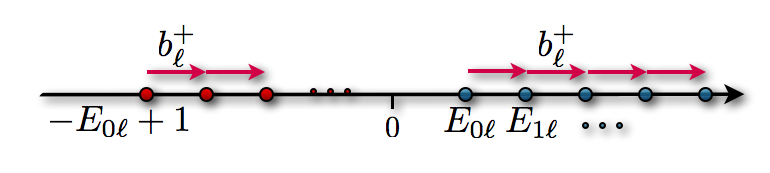}
\caption{\small{Spectrum of the radial oscillator Hamiltonian $H_\ell$. The blue circles represent the physical solutions starting from $E_0$ and the red ones the non-physical solutions departing from $-E_{0\ell}+1$.}}\label{speRO}
\end{figure}

Moreover, the change $\ell\rightarrow -(\ell+1)$ produces the other two factorizations, although this causes small changes in the equations. For the factorizations we have,
\begin{equation}
H_\ell=a_{-(\ell+1)}^- a_{-(\ell+1)}^+-\frac{\ell}{2}-\frac{3}{4}=a_{-\ell}^+ a_{-\ell}^- -\frac{\ell}{2}+\frac{1}{4},
\end{equation}
for the intertwinings
\begin{equation}
H_{\ell-1} a_{-\ell}^- = a_{-\ell}^-\left(H_{\ell}-\frac{1}{2}\right),\qquad
H_{\ell} a_{-\ell}^+ = a_{-\ell}^+\left(H_{\ell-1}+\frac{1}{2}\right),
\end{equation}
and for the eigenvalue equations
\begin{equation}
H_{\ell-1} (a_{-\ell}^- \psi_{n\ell}) = \left(E_{n\ell}- \frac{1}{2}\right)(a_{-\ell}^-\psi_{n\ell}),\ \
H_{\ell+1} (a_{-(\ell+1)}^+ \psi_{n\ell}) = \left(E_{n\ell}+\frac{1}{2}\right)(a_{-(\ell+1)}^+\psi_{n\ell}).
\end{equation}
Neither $a_\ell^\pm$ or $a_{-\ell}^\pm$ are ladder operators; nevertheless, through them we can define second-order ones. From diagram in figure~\ref{diaRO} we can see that the joint action of two appropriate shift operators lead to an effective ladder operator. Indeed, let us take $b_\ell^\pm$ such that
\begin{equation}
b_\ell^- = a_{-(\ell+1)}^-a_{\ell+1}^-=a_\ell^-a_{-\ell}^-,\qquad
b_\ell^+ = a_{\ell+1}^+a_{-(\ell+1)}^+=a_{-\ell}^+a_{\ell}^+.
\end{equation}
Then we can easily show that 
\begin{equation}
H_\ell b_\ell^- = b_\ell^-(H_\ell-1), \qquad H_\ell b_\ell^+ = b_\ell^+(H_\ell + 1),
\end{equation}
i.e., the following commutators are obeyed
\begin{equation}
[H_\ell,b_\ell^{\pm}]=\pm b_\ell^\pm.
\end{equation}
This proves that $b_\ell^\pm$ are ladder operators of $H_\ell$. Their explicit form is
\begin{equation}
b_\ell^\pm=\frac{1}{2}\left(\frac{\text{d}^2}{\text{d}x^2}\mp x\frac{\text{d}}{\text{d}x}+\frac{x^2}{4}-\frac{\ell(\ell+1)}{x^2}\mp\frac{1}{2}\right).
\end{equation}
We can obtain the eigenstates of $H_\ell$ if we start from the {\it ground state} $\psi_{0\ell}$, an eigenstate of $H_\ell$ such that $b_\ell^- \psi_{0\ell}=0$. In this systems there are two such a states,
\begin{subequations}
\begin{alignat}{3}
\psi_{\mathcal{E}_1} &\propto x^{\ell+1}\exp(-x^2/4), & \qquad \mathcal{E}_1 & =\frac{\ell}{2}+\frac{3}{4}\equiv E_{0\ell},\\
\psi_{\mathcal{E}_2} &\propto x^{-\ell}\exp(-x^2/4), & \qquad \mathcal{E}_2 & = -\frac{\ell}{2}+\frac{1}{4}= -E_{0\ell}+1,
\end{alignat}\label{solsRO}
\end{subequations}
\hspace{-1.5mm}but only the first one fulfills the boundary conditions and therefore leads to a ladder of {\it physical} eigenfunctions. The spectrum of the radial oscillator is therefore
\begin{equation}
\text{Sp}(H_\ell)=\{E_{n\ell}=n+\frac{\ell}{2}+\frac{3}{4}, n=0,1,\dots \}.
\end{equation}
We can see a diagram of this spectrum in figure~\ref{speRO} where we represent both the physical and non-physical solutions obtained from the extremal states of equations~\eqref{solsRO}.

An analogue of the number operator can now be defined for the radial oscillator as
\begin{equation}
b_\ell^+b_\ell^-=(H_\ell-\mathcal{E}_1)(H_\ell-\mathcal{E}_2)=\left(H_\ell-\frac{\ell}{2}-\frac{3}{4}\right)\left(H_\ell+\frac{\ell}{2}-\frac{1}{4}\right).
\end{equation}

In order to perform the SUSY transformations, we need to find the general solution of the Schr\"odinger equation for any factorization energy $\epsilon$, which is given by \cite{CFNN04,JR98,Car01}
\begin{align}
u(x)=\, & x^{-\ell}\text{e}^{-x^2/4}\left[{}_1F_1\left(\frac{1-2\ell-4\epsilon}{4},\frac{1-2\ell}{2};\frac{x^2}{2}\right)\right. \nonumber\\
& + \left. \nu \frac{\Gamma\left(\frac{3+2\ell-4\epsilon}{4}\right)}{\Gamma\left(\frac{3+2\ell}{2}\right)}\left(\frac{x^2}{2}\right)^{\ell+1/2}{}_1F_1\left(\frac{3+2\ell-4\epsilon}{4},\frac{3+2\ell}{2};\frac{x^2}{2}\right)\right].\label{solRO}
\end{align}
Thus, the first-order SUSY partner potential of the radial oscillator becomes
\begin{equation}
V_1(x)=\frac{x^2}{8}+\frac{\ell(\ell+1)}{2x^2}-[\ln u(x)]''.
\end{equation}
Three conditions must be fulfilled to avoid singularities in this transformation
\begin{equation}
x>0, \quad \epsilon<E_{0\ell}, \quad \nu\geq -\frac{\Gamma\left(\frac{1-2\ell}{2}\right)}{\Gamma\left(\frac{1-2\ell-4\epsilon}{4}\right)}.\label{condRO}
\end{equation}

\begin{figure}\centering
\includegraphics[scale=0.335]{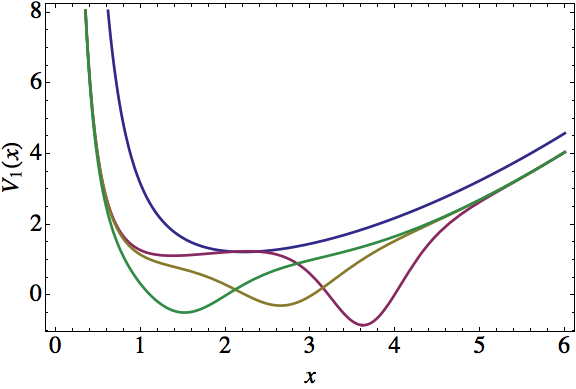}
\includegraphics[scale=0.35]{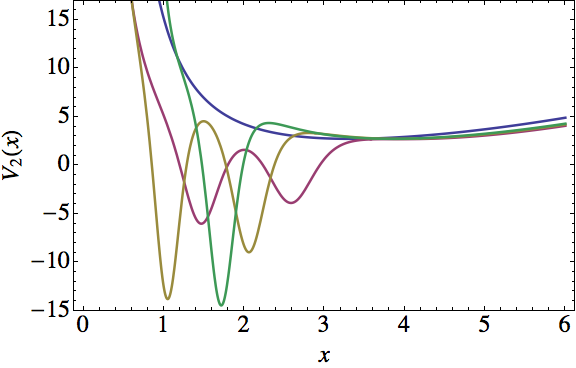}
\caption{\small{SUSY partner potentials of the radial oscillator (blue). The left plot is for $k=1$, $\ell=2$, $\epsilon=1/2$, and $\nu=\{-$0.59 (magenta), $-0.4$ (yellow), 1 (green)$\}$. The right plot is for $k=2$, $\ell=5$, $\nu_1=1$, and $\epsilon_1=\{$0 (magenta), $-2$ (yellow), $-4$ (green)$\}$, where we have taken $u_2=b_\ell^- u_1$ and thus $\epsilon_2=\epsilon_1 - 1$.}}\label{potRO}
\end{figure}

Let us  apply now the $k$th-order SUSY QM by taking $k$ appropriate solutions $\{u_k,\dots,u_1\}$, as given in equation (\ref{solRO}), associated to the factorization energies 
$\epsilon_k < \epsilon_{k-1} < \cdots < \epsilon_1 < E_0$. The SUSY deformed potential is now given by
\begin{equation}
V_k(x)=\frac{x^2}{8}+\frac{\ell(\ell+1)}{2x^2}-\{\ln[W(u_1,\dots , u_k)]\}'',
\end{equation}
with the spectrum $\text{Sp}(H_k)=\{\epsilon_k,\dots,\epsilon_1,E_{0\ell},E_{1\ell},\dots \}$. In figure~\ref{potRO} we show some examples of first- and second-order SUSY partner potentials of the radial oscillator. 

As for the harmonic oscillator, we can define again a natural pair of ladder operators $L_k^{\pm}$ for the $k$-SUSY partners $H_k$ of the radial oscillator as $L_k^\pm=B_k^+b_\ell^\pm B_k^-$, which are of $(2k+2)$th-order and fulfill $[H_k,L_k^\pm]=\pm L_k^\pm$.

From the intertwining relations the analogue of the number operator for the radial oscillator is obtained
\begin{equation}
N(H_k)\!=\!L_k^+L_k^-\!=\!\!\prod_{j=1}^{2k+2}\!(H_k-\mathcal{E}_j)
		  \!=\!\left(\!\!H_k-\frac{\ell}{2}-\frac{3}{4}\!\right)\left(\!\!H_k+\frac{\ell}{2}-\frac{1}{4}\!\right)\!\!\prod_{j=1}^{k}\!(H_k-\epsilon_j)(H_k-\epsilon_j-1).
\end{equation}

\section{Painlev\'e equations}\label{peqs}
The special functions play an important role in the study of linear differential equations, which are also of great importance in mathematical physics. Examples of these are Airy $Ai(z)$, Bessel $J_{\nu}(z)$, parabolic cylindrical $D_{\nu}(z)$, Whittaker $M_{\kappa,\mu}(z)$, confluent hypergeometric $_1F_1(a,b;z)$ and hypergeometric functions 
$_2F_1(a,b,c;z)$. Some of them are solutions of linear ordinary differential equations with rational coefficients which receive the same name as the functions. For example, the Bessel functions are solutions of Bessel equation, the simplest second-order linear differential equation with one irregular singularity.

Painlev\'e equations play an analogous role for non-linear differential equations. In fact some specialists \cite{IKSY91,CM08} consider that during the 21st century, Painlev\'e functions will be new members of the special functions. The corresponding equations are non-linear second-order ordinary differential equations that were found by Painlev\'e and others at the beginning of the 20th century by purely mathematical reasons, which are denoted by PI,...,PVI. Our interest in these lecture notes is focused in PIV and PV equations. Note that PIV equation appears in the asymptotic behaviour of non-linear evolution equations \cite{SA81}, correlation functions of the XY model \cite{STN01}, bidimensional Ising model \cite{MPS02}, Einstein axialsymmetric equations \cite{GMS08}, negative curvature surfaces \cite{CX10}, among others. On the other hand, PV equation arises in the study of correlation functions in condense matter \cite{Kan02}, in Maxwell-Bloch systems in electrodynamics \cite{Win92}, and in the symmetry reduction for the stimulated Raman scattering \cite{Lev92}.

The ideas of Paul Painlev\'e allowed to distinguish six families of non-linear second-order differential equations, traditionally represented by \cite{IKSY91,Sac91}
\begin{subequations}
\begin{align}
\text{PI}\ :\ & w'' = 6 w^2 + z, \\
\text{P{II}} \ :\ & w'' = 2 w^3 + zw + a , \\
\text{P{III}} \ :\ & w'' = \frac{1}{w}{w'}^2  - \frac{1}{z}w' + \frac{1}{z}(a w^2 + b) + c w^3 + \frac{d}{w}, \\
\text{P{IV}} \ :\ & w'' = \frac{1}{2w}{w'}^2  + \frac{3}{2}w^3 + 4 z w^2 + 2(z^2 - a)w + \frac{b}{w},\label{PIVlib}\\ 
\text{P{V}} \ :\ &  
w'' = \left[\frac{1}{2w} + \frac{1}{w - 1}\right]{w'}^2 - \frac{1}{z}w'  + \frac{(w - 1)^2}{z^2}\left[a w + \frac{b}{w}\right] \nonumber\\
\phantom{.} &\hspace{10mm} 
+ \frac{c w}{z} + \frac{d w(w + 1)}{w - 1},\\
\text{P{VI}} \ :\ & w'' =  \frac12 \left[\frac{1}{w} + \frac{1}{w -1} + \frac{1}{w - z}\right]{w'}^2 - \left[\frac{1}{z} + \frac{1}{z - 1} + \frac{1}{w - z}\right]w'  \nonumber\\
\phantom{.} & \hspace{9mm} + \frac{w(w-1)(w-z)}{z^2(z-1)^2}\left[a + \frac{b z}{w^2} + \frac{c(z -1)}{(w-1)^2} + \frac{d z(z -1)}{(w - z)^2}\right] ,
\end{align}\label{painleves}
\end{subequations}
\hspace{-1mm}where $a,b,c,d\in\mathbb{C}$ are constants.

For arbitrary values of the parameters $a,b,c,d$, the general solutions of the Painlev\'e equations are {\it transcendental}, i.e., they cannot be expressed in closed form in terms of elementary functions. However, for specific values of the parameters they have special solutions in terms of elementary or special functions \cite{BCH95}. We know the fundamental role that second-order differential equations play in the description of physical phenomena, in contrast to higher-order equations. Nevertheless, until recently, scientists were not aware of any non-linear ordinary differential equation without essential singularities that have a physical significance.

\section{Polynomial Heisenberg algebras}\label{pha}

The polynomial Heisenberg algebras (PHA) are deformations of the Heisenberg-Weyl algebra for which the commutators of the Hamiltonian $H$ with the ladder operators $\mathcal{L}_m^{\pm}$ are the same as for the harmonic oscillator, 
\begin{equation}
[H,\mathcal{L}_m^{\pm}]=\pm \mathcal{L}_m^{\pm}, \label{pha1}
\end{equation}
while the deformation is contained in the following commutator:
\begin{equation}
[\mathcal{L}_m^{-},\mathcal{L}_m^{+}]\equiv N_m(H+1) - N_m(H)= P_{m-1} (H),\label{pha2}
\end{equation}
where $\mathcal{L}_m^{\pm}$ are $m$th-order differential ladder operators, $P_{m-1}(H)$ is a $(m-1)$th-order polynomial of $H$ and $N_m(H)\equiv \mathcal{L}_m^{+}\mathcal{L}_m^{-}$ is a $m$th-order polynomial in $H$, which is the analogous of the number operator for the harmonic oscillator and is factorized as
\begin{equation}
N_m(H)=\prod_{i=1}^{m}(H-\mathcal{E}_{i}),\label{facN}
\end{equation}
$\mathcal{E}_i$ being the energies associated with the {\it extremal states}. Taking into account the degree of the polynomial $P_{m-1} (H)$ in (\ref{pha2}) we will say that this is a PHA of $(m-1)$th-order.

\begin{figure}
\centering
\includegraphics[scale=0.24]{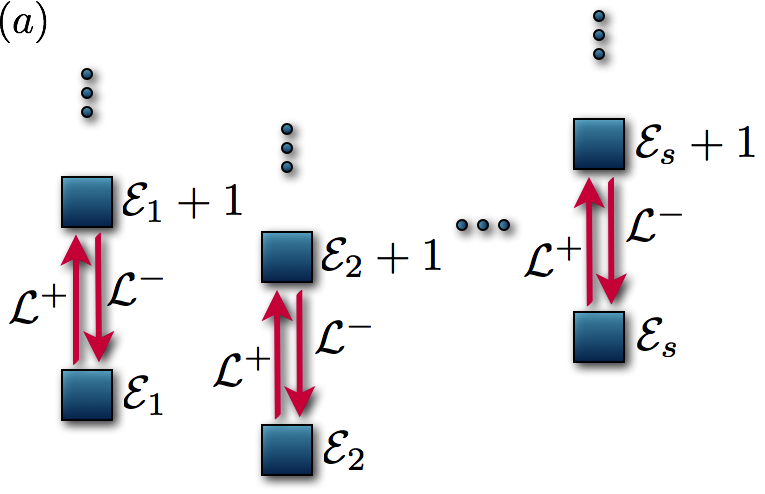}\hspace{10mm}
\includegraphics[scale=0.24]{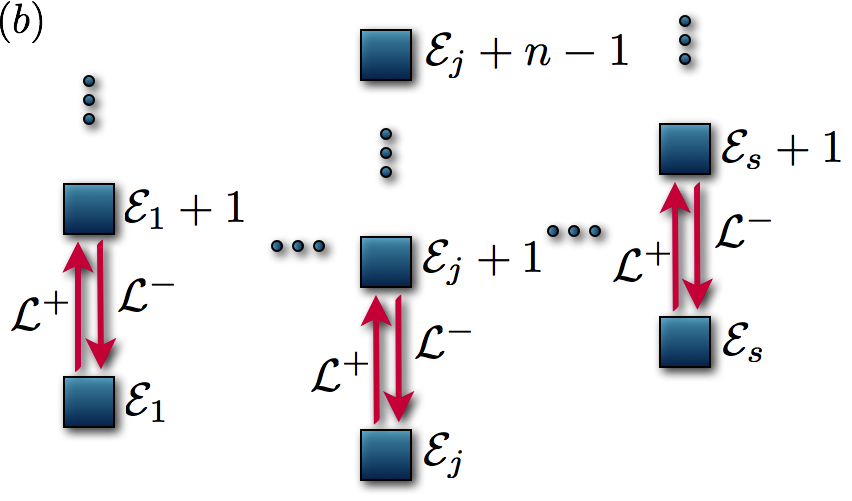}
\caption{\small{In (a) we show the spectrum for a Hamiltonian with $s$ physical extremal states. In general, each one of them has associated one infinite ladder. In (b) we show a spectrum where $\psi_{\mathcal{E}_j}$ fulfill condition \eqref{finita} and therefore the system has $s-1$ infinite and one finite (the $j$-th) ladders.}}\label{fig.pha1b}
\end{figure}

The algebraic structure generated by $\{H,\mathcal{L}_m^{-},\mathcal{L}_m^{+}\}$ provides information about the spectrum of $H$, $\text{Sp}(H)$ \cite{DEK92,FH99,ACIN00}. In fact, let us consider the $m$th-dimensional solution space of the differential equation $\mathcal{L}_m^{-}\psi=0$, called the {\it kernel} of $\mathcal{L}_m^{-}$ and denoted as $\mathcal{K}_{\mathcal{L}_m^{-}}$. Then
\begin{equation}
\mathcal{L}_m^{+}\mathcal{L}_m^{-}\psi = \prod_{i=1}^{m}(H-\mathcal{E}_i )\psi=0.
\end{equation}
Since $\mathcal{K}_{\mathcal{L}_m^{-}}$ is invariant under $H$, then it is natural to select the corresponding eigenfunctions of $H$ as basis for the solution space, i.e., $H\psi_{\mathcal{E}_i}=\mathcal{E}_i \psi_{\mathcal{E}_i}$. Therefore, $\psi_{\mathcal{E}_i}$ are the extremal states of $m$ mathematical ladders with spacing $\Delta E=1$ that start from $\mathcal{E}_i$. Let $s$ be the number of those states with physical significance, $\{ \psi_{\mathcal{E}_i}, i=1,\dots ,s \}$; then, using $\mathcal{L}_m^{+}$ we can construct $s$ physical energy ladders on infinite length, as shown in figure \ref{fig.pha1b}(a).

It is possible that for a ladder starting from $\mathcal{E}_j$ there exists an integer $n\in N$ such that
\begin{equation}
(\mathcal{L}_m^{+})^{n-1}\psi_{\mathcal{E}_j} \neq 0,\qquad
(\mathcal{L}_m^{+})^{n}\psi_{\mathcal{E}_j} = 0.
\label{finita}
\end{equation}
Then, if we analize $\mathcal{L}_m^{-}(\mathcal{L}_m^{+})^{n}\psi_{\mathcal{E}_j} = 0$ it is seen that other root of equation~\eqref{facN} must fulfill $\mathcal{E}_k = \mathcal{E}_j +n$, where $k\in \{s+1,\dots ,m\}$ and $\ j\in \{1,\dots,s \}$. Therefore, Sp$(H)$ contains $s-1$ infinite ladders and a finite one of length $n$, that starts in $\mathcal{E}_j$ and finish in $\mathcal{E}_j +n-1$ (see figure \ref{fig.pha1b}(b)).

We conclude that the spectrum of systems described by an $(m-1)$th-order PHA can have at most $m$ infinite ladders. Note that the annihilation and creation operators of the harmonic oscillator $a^\pm$, together with the Hamiltonian, satisfy equations (\ref{pha1}--\ref{facN}). Moreover, higher-order PHA with odd $m$ can be constructed simply by taking $\mathcal{L}_m^{-}=a^{-}\mathcal{P}(H)$, $\mathcal{L}_m^{+}=\mathcal{P}(H)a^{+}$, where $\mathcal{P}(H)$ is a real polynomial of $H$ \cite{DGRS99}. These deformations are called \emph{reducible}, and in this context they are somehow artificial since our system already has operators $a^\pm$ that fulfill a lower-order algebra.

\section{General systems ruled by PHA}
\label{secpha3}
It is important to identify the general systems ruled by PHA, characterized by a Schr\"odinger Hamiltonian:
\begin{equation}
H = - \frac12 \frac{\text{d}^2}{\text{d}x^2}+ V(x).
\end{equation}
We will see in this section that the difficulties in the analysis grow with the order of the PHA: for zeroth and first order the systems become the harmonic and radial oscillators respectively \cite{Adl93,Fer84D,DEK92,SRK97}. On the other hand, for second and third order PHA, the determination of the potentials  reduces to find solutions of Painlev\'e IV and V equations, respectively \cite{Adl93,WH03}.
\subsection{Zeroth-order PHA: first-order ladder operators.}
Let us start by taking the first-order ladder operators $\mathcal{L}_1^{\pm}$ in the way
\begin{equation}
\mathcal{L}_1^+=\frac{1}{2^{1/2}}\left[-\frac{\text{d}}{\text{d}x}+f(x)\right],\quad \mathcal{L}_1^-=(\mathcal{L}_1^+)^\dag,
\end{equation}
which satisfy equation \eqref{pha1}. Thus, a system involving $V$, $f$, and their derivatives is obtained
\begin{equation}
f' - 1 = 0,\qquad V' - f = 0.
\end{equation}

Up to coordinate and energy displacements, it turns out that $f(x) = x$ and $V(x) = x^2/2$. This potential has one equidistant infinite ladder starting from the extremal state $\psi_{\mathcal{E}_1}=\pi^{-1/4} \exp(-x^2/2)$, which is a normalized eigenfunction of $H$ with eigenvalue $\mathcal{E}_1 = 1/2$ annihilated by $\mathcal{L}_1^-$. Here, the number operator is linear in $H$, $N_1(H) = H - \mathcal{E}_1$, i.e., the most general system obeying the zeroth-order PHA of section \ref{pha} is the harmonic oscillator. Its natural ladder operators are the standard first-order annihilation and creation operators $\mathcal{L}_1^\pm=a^\pm$. They generate the Heisenberg-Weyl algebra, which has been widely studied.

\subsection{First-order PHA: second-order ladder operators.}
Let us suppose now that
\begin{equation}
\mathcal{L}_2^+ = \frac{1}{2}\left[\frac{\text{d}^2}{\text{d}x^2} + g(x)\frac{\text{d}}{\text{d}x} + h(x)\right],\quad \mathcal{L}_2^- = (\mathcal{L}_2^+)^\dag.
\end{equation}
Then, equation \eqref{pha1} leads to a system of equations for $V$, $g$, $h$, and their derivatives
\begin{equation}
g' + 1 = 0,\qquad
h' + 2V' + g = 0,\qquad
h'' +2V'' +2gV' +2h=0.
\end{equation}
The general solution (up to coordinate and energy displacements) is given by
\begin{equation}
g(x)=-x,\qquad
h(x)= \frac{x^2}{4} - \frac{\gamma}{x^2} - \frac{1}{2},\qquad
V(x)= \frac{x^2}{8} +\frac{\gamma}{2x^2}, \label{m1sols}
\end{equation}
\hspace{-1mm}where $\gamma$ is a real constant. The potential of equation~\eqref{m1sols} has two equidistant energy ladders (not necessarily physical), generated by acting with powers of $\mathcal{L}_2^+$ on the two extremal states
\begin{equation}
\psi_{\mathcal{E}_1}\propto x^{1/2+\sqrt{\gamma+1/4}} \exp\left(-\frac{x^2}{4}\right),\qquad
\psi_{\mathcal{E}_2}\propto x^{1/2-\sqrt{\gamma+1/4}} \exp\left(-\frac{x^2}{4}\right).
\end{equation}

Let us recall that $\mathcal{L}_2^- \psi_{\mathcal{E}_j}=0$ and $(H-\mathcal{E}_j)\psi_{\mathcal{E}_j}=0$, where
\begin{equation}
\mathcal{E}_1=\frac{1}{2}+\frac{1}{2}\sqrt{\gamma+\frac{1}{4}},\qquad
\mathcal{E}_2=\frac{1}{2}-\frac{1}{2}\sqrt{\gamma+\frac{1}{4}}.
\end{equation}
Now $N_2(H)$ is quadratic in $H$, i.e., $N_2(H)=(H-\mathcal{E}_1)(H-\mathcal{E}_2)$. The potentials can be expressed as
\begin{equation}
V(x)= \frac{x^2}{8} + \frac{\ell(\ell+1)}{2x^2},\qquad x>0, \quad \ell\geq 0,
\end{equation}
which were obtained by making $\gamma = \ell(\ell+1),$ $\ell \geq 0$. Thus, the general systems having second-order ladder operators are described by the radial oscillator potentials. The natural ladder operators for the first-order PHA are the second-order ones of the radial oscillator, $\mathcal{L}_2^\pm=b_\ell^\pm \equiv b^\pm$, which generate the $\text{so}(2,1)$ algebra.

\subsection{Second-order PHA: third-order ladder operators.}\label{secondPHA}
\begin{figure}
\begin{center}
\includegraphics[scale=0.26]{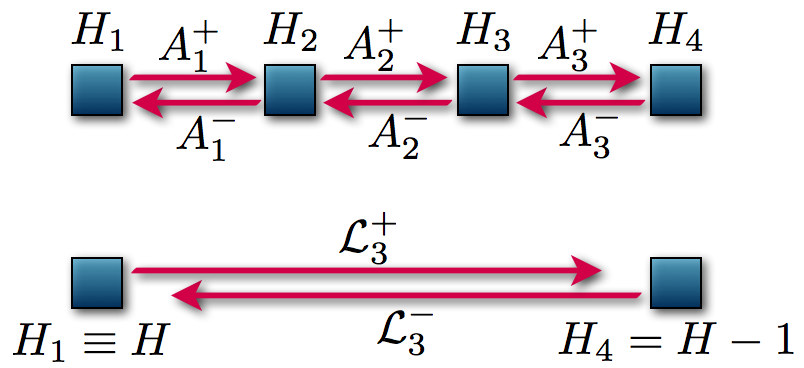}
\end{center}
\vspace{-5mm}
\caption{\small{Diagram of the two globally equivalent SUSY transformations. Above: the three-step first-order operators $A_1^{\pm}$, $A_2^{\pm}$, and $A_3^{\pm}$ allow to accomplish the transformation. Below: the direct transformation achieved through the third-order operators $\mathcal{L}_3^{\pm}$.}}
\label{diasusy}
\end{figure}

In this case, $\mathcal{L}_3^{\pm}$ will be third-order differential ladder operators. Now, we propose a closed-chain of three first-order SUSY transformations \cite{VS93,DEK94,ACIN00,Adl94,Gra04,Mar09b,mq13} so that 
\begin{subequations}
\begin{align}
\mathcal{L}_3^{+}&=A_{3}^{+}A_{2}^{+}A_{1}^{+}=
\frac{1}{2^{3/2}}\left(\frac{\text{d}}{\text{d}x}-f_3\right)\left(\frac{\text{d}}{\text{d}x}-f_2\right)\left(\frac{\text{d}}{\text{d}x}-f_1\right),\\
\mathcal{L}_3^{-}&=A_{1}^{-}A_{2}^{-}A_{3}^{-}=
\frac{1}{2^{3/2}}\left(-\frac{\text{d}}{\text{d}x}-f_1\right)\left(-\frac{\text{d}}{\text{d}x}-f_2\right)\left(-\frac{\text{d}}{\text{d}x}-f_3\right)\label{lmenos}.
\end{align}\label{eles}
\end{subequations}
\hspace{-2mm}In general, $(\mathcal{L}_3^{-})^{\dag}\neq \mathcal{L}_3^{+}$, except in the case where all $f_j \in \mathbb{R}$. The pair $A_j^\pm$ fulfill three intertwining relations of kind
\begin{equation}
H_{j+1}A^{+}_j =A^{+}_{j}H_{j},\quad
H_{j}A^{-}_j =A^{-}_{j}H_{j+1},\label{fac2}
\end{equation}
where $j=1,2,3$. In figure~\ref{diasusy} we present a diagram of the intertwining relations.

If we equate the two different factorizations associated with each $H_j$ in equation \eqref{fac2} which lead to the same Hamiltonians, we get
\begin{equation}
H_{j}=A_{j}^{-}A_{j}^{+}+\epsilon_{j},\quad H_{j+1}=A_{j}^{+}A_{j}^{-}+\epsilon_{j},\quad j=1,2,3.\label{facs2pha}
\end{equation}
In addition, the closure condition is given by $H_4=H_1-1\equiv H-1$. By making the corresponding operator products we get the following system of equations \cite{VS93,MN08,Adl94}
\begin{subequations}
\begin{align}
f_1'+f_2'&=f_1^2-f_2^2+2(\epsilon_1-\epsilon_2),\label{f1}\\
f_2'+f_3'&=f_2^2-f_3^2+2(\epsilon_2-\epsilon_3),\label{f2}\\
f_3'+f_1'&=f_3^2-f_1^2+2(\epsilon_3-\epsilon_1+1).\label{f3}
\end{align}
\end{subequations}

Eliminating $f_2^2$ from equations \eqref{f1} and \eqref{f2} we get
\begin{equation}
f_1'+2f_2'+f_3'=f_1^2-f_3^2+2(\epsilon_1-\epsilon_3),
\end{equation}
and from here we substitute $f_3^2$ from equation \eqref{f3} to obtain $f_1'+f_2'+f_3'=1$, which, after integration becomes
\begin{equation}
f_1+f_2+f_3=x.\label{f4}
\end{equation}
Now, substituting equation~\eqref{f4} into \eqref{f1} to eliminate $f_2$
\begin{equation}
f_1=\frac{x-f_3}{2}+\frac{1-f_3'}{2(x-f_3)}-\frac{\epsilon_1-\epsilon_2}{x-f_3}.
\end{equation}
Let us define now a useful new function as $g \equiv f_3-x$, from which we get
\begin{equation}
f_1=-\frac{g}{2}+\frac{g'}{2g}+\frac{\epsilon_1-\epsilon_2}{g}.\label{f1b}
\end{equation}
Similarly, by plugging equation~\eqref{f4} into \eqref{f1} to eliminate $f_1$ and using $g$ it turns out that
\begin{equation}
f_2=-\frac{g}{2}-\frac{g'}{2g}-\frac{\epsilon_1-\epsilon_2}{g}.\label{f2b}
\end{equation}
Now that we have $f_1,f_2,f_3$ in terms of $g$, we replace them in equation~\eqref{f3} to obtain
\begin{equation}
gg'' = \frac{1}{2}(g')^2 + \frac{3}{2}g^4 + 4g^3x+ 2g^2\left(x^2-a\right)+b,\label{PIV}
\end{equation}
with parameters
\begin{equation}
a=\epsilon_1+\epsilon_2-2\epsilon_3 -1,\quad b=-2(\epsilon_1-\epsilon_2)^2,\label{abe}
\end{equation}
which is the Painlev\'e IV (PIV) equation \cite{IKSY91,VS93,Adl94,BCH95,ACIN00,WH03} (compare with \eqref{PIVlib}). Since, in general $f\in\mathbb{C}$ then $g\in\mathbb{C}$. In addition, $\epsilon_i\in\mathbb{C}$ which implies that $a,\, b\in\mathbb{C}$ and therefore $g$ is a complex solution to the PIV equation associated with the complex parameters $a,\, b$.

With the solution $g(x)$ of (\ref{PIV}) one can find the new potential $V(x)$ as
\begin{equation}
V(x)=\frac{x^2}{2}-\frac{g'}{2}+\frac{g^2}{2}+xg+\epsilon_3+\frac{1}{2}.\label{VPIV}
\end{equation}
Now, the energies of the extremal states are defined as the roots of the generalized number operator, which is cubic in this case
\begin{equation}
N_3(H)=(H-\mathcal{E}_1)(H-\mathcal{E}_2)(H-\mathcal{E}_3).\label{q3}
\end{equation}
Using the definitions from equations~(\ref{eles}--\ref{facs2pha}) we thus have $\mathcal{E}_i=\epsilon_i+1,\ i=1,2,3$.

As can be seen, if one solution $g(x)$ of the PIV equation is obtained for certain values of ${\cal E}_1, \ {\cal E}_2, \ {\cal E}_3$, then $V(x)$ and $\mathcal{L}_3^\pm$ are completely determined (see equations \eqref{f1b}, \eqref{f2b} and \eqref{VPIV}). Moreover, the three extremal states are obtained from $\mathcal{L}_3^{-} \psi_{\mathcal{E}_j}=(H-\mathcal{E}_j)\psi_{\mathcal{E}_j}=0$, $j=1,2,3$, which leads to 
\begin{subequations}
\begin{align}
 \psi_{{\cal E}_1} &\propto \left( \frac{g'}{2g} - \frac{g}{2} - \frac{1}{g}\sqrt{-\frac{b}{2}} - x\right)
\exp\left[\int\left( \frac{g'}{2g} + \frac{g}{2} - \frac{1}{g}\sqrt{-\frac{b}{2}} \right) \text{d}x \right], \label{exes1} \\
\psi_{{\cal E}_2} &\propto \left( \frac{g'}{2g} - \frac{g}{2} + \frac{1}{g}\sqrt{-\frac{b}{2}} - x\right)
\exp\left[\int\left( \frac{g'}{2g} + \frac{g}{2} + \frac{1}{g}\sqrt{-\frac{b}{2}} \right) \text{d}x \right], \label{exes2} \\
\psi_{{\cal E}_3} &\propto \exp\left( - \frac{x^2}{2} - \int g\, \text{d}x\right). \label{exes3}
\end{align}\label{exes}
\end{subequations}
\hspace{-1.8mm}The corresponding physical ladders of our system are obtained departing from the physically admissible extremal states. In this way we can determine the spectrum of $H$.

On the other hand, if a system with third-order differential ladder operators is found, it is possible to design a method to obtain solutions of the PIV equation. The key point is to identify the extremal states of our system; then, from equation~\eqref{exes3} it is easy to see that
\begin{equation}
g(x) = - x - \{\ln[\psi_{{\cal E}_3}(x)]\}'.\label{exes3des}
\end{equation}
Note that, if we permute cyclically the indices assigned to the extremal states $\psi_{{\cal E}_1},\, \psi_{{\cal E}_2},\, \psi_{{\cal E}_3}$, we will obtain three solutions of the PIV equation with different parameters $a,b$.

Hence, we have found a recipe for building systems ruled by second-order PHA, defined by equations~(\ref{pha1}--\ref{facN}): first find a function $g(x)$ that solves the PIV equation~\eqref{PIV}; then calculate the potential using equation~\eqref{VPIV}, and its three ladders from the extremal states given by equations~\eqref{exes}.

\subsection{Third-order PHA: fourth-order ladder operators.}
\begin{figure}
\begin{center}
\includegraphics[scale=0.26]{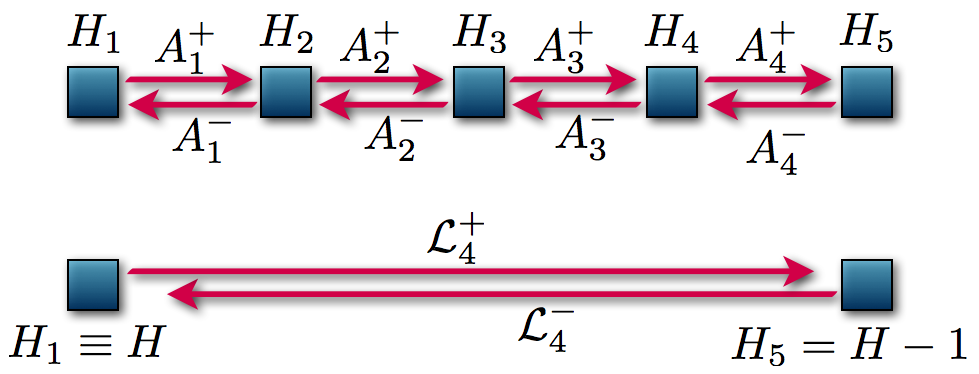}
\end{center}
\vspace{-5mm}
\caption{\small{Diagram representing the two globally equivalent SUSY transformations. Above: the four-step first-order operators $A_1^{\pm}$, $A_2^{\pm}$, $A_3^{\pm}$ and $A_4^\pm$. Below: the direct transformation achieved through the fourth-order operators $\mathcal{L}_4^{\pm}$.}}
\label{diasusyRO}
\end{figure}

Now $\mathcal{L}_4^{\pm}$ will be fourth-order ladder operators. We propose again a closed-chain as follows \cite{Adl94}
\begin{subequations}
\begin{align}
\mathcal{L}_4^{+}&=A_4^+A_{3}^{+}A_{2}^{+}A_{1}^{+}=
\frac{1}{2^{2}}\left(\frac{\text{d}}{\text{d}x}-f_4\right)\left(\frac{\text{d}}{\text{d}x}-f_3\right)\left(\frac{\text{d}}{\text{d}x}-f_2\right)\left(\frac{\text{d}}{\text{d}x}-f_1\right),\\
\mathcal{L}_4^{-}&=A_{1}^{-}A_{2}^{-}A_{3}^{-}A_4^-=
\frac{1}{2^{2}}\left(-\frac{\text{d}}{\text{d}x}-f_1\right)\left(-\frac{\text{d}}{\text{d}x}-f_2\right)\left(-\frac{\text{d}}{\text{d}x}-f_3\right)\left(-\frac{\text{d}}{\text{d}x}-f_4\right)\label{lmenosRO}.
\end{align}\label{elesRO}
\end{subequations}
\hspace{-1mm}Each pair of operators $A_j^-$, $A_j^+$ intertwines two Hamiltonians $H_j$ and $H_{j+1}$ in the way
\begin{equation}
H_{j+1}A^{+}_j =A^{+}_{j}H_{j},\qquad
H_{j}A^{-}_j =A^{-}_{j}H_{j+1},
\end{equation}
where $j=1,2,3,4$. This leads to the following factorizations of the Hamiltonians
\begin{equation}
H_{j}=A_{j}^{-}A_{j}^{+}+\epsilon_{j},\qquad H_{j+1}=A_{j}^{+}A_{j}^{-}+\epsilon_{j},\qquad j=1,2,3,4.\label{cloconRO}
\end{equation}
To accomplish the closed-chain we need the closure condition given by $H_5=H_1-1\equiv H-1$. In figure~\ref{diasusyRO} we show a diagram representing the transformation and the closure relation. By making the corresponding operator products we obtain the following systems of equations
\begin{subequations}
\begin{align}
f_1'+f_2'=f_1^2-f_2^2+2(\epsilon_1-\epsilon_2),\qquad
& f_2'+f_3'=f_2^2-f_3^2+2(\epsilon_2-\epsilon_3),\label{f2RO}\\
f_3'+f_4'=f_3^2-f_4^2+2(\epsilon_3-\epsilon_4),\qquad
& f_4'+f_1'=f_4^2-f_1^2+2(\epsilon_4-\epsilon_1+1).\label{f3RO}
\end{align}\label{efesRO}
\end{subequations}
\hspace{-1mm}Up to now, the method employed for this case is very similar to the one for the second-order PHA, and indeed can be taken as its generalization; nevertheless, the similarity ends now.

Let us simplify the notation making $\alpha_1=\epsilon_1-\epsilon_2$, $\alpha_2=\epsilon_2-\epsilon_3$, $\alpha_3=\epsilon_3-\epsilon_4$, and $\alpha_4=\epsilon_4-\epsilon_1+1$. If we sum all equations~\eqref{efesRO} we obtain
\begin{equation}
f_1+f_2+f_3+f_4=x.\label{sumas}
\end{equation}
Since the system is over-determined, we can use a constrain $A$ as
\begin{equation}
f_1^2-f_2^2+f_3^2-f_4^2=\alpha_4-\alpha_3+\alpha_2-\alpha_1\equiv A.\label{restr1}
\end{equation}
Using \eqref{sumas} and \eqref{restr1} we can reduce the system of equations \eqref{efesRO} to a second-order one. Let us denote $g\equiv -f_1-f_2$, $p\equiv f_1-f_2$, $q\equiv f_2+f_3$. Then equations~\eqref{f2RO} are written as
\begin{equation}
g'=gp-2\alpha_1,\quad q' =-q(q+g+p)+2\alpha_2,\label{gqp}
\end{equation}
and the restriction $A$ is expressed as
\begin{equation}
x p + (g+x)(2q-x)=A.\label{restr2}
\end{equation}
Now we have the system of three equations \eqref{gqp} and \eqref{restr2}. We define $t\equiv 2q-x$ and then we clear $p$ from equation~\eqref{restr2}
\begin{equation}
p=\frac{1}{x}[A-(x+g)t].
\end{equation}
Then we substitute this into both equations~\eqref{gqp} to obtain a two-equation system
\begin{equation}
g' =\frac{g}{x}[A-(x+g)t]-2\alpha_1,\qquad t'=(t+x)\left(\frac{gt-A}{x}+\frac{t-x}{2}-g\right)+4\alpha_2-1.\label{gh}
\end{equation}

Now, let us define two new functions $w$ and $v$ as
\begin{equation}
xt(x)  = v(x^2),\quad g(x)  = \frac{x}{w(x^2)-1}, \label{vw}
\end{equation}
and we change $x^2\rightarrow z$, which takes the system to
\begin{subequations}
\begin{align}
v' & = \left(\frac{v^2}{4z}-\frac{z}{4}\right)\left(\frac{w+1}{w-1}\right)+(1-A)\frac{v}{2z}+2\alpha_2-\frac{A}2 - \frac12,\label{vw1}\\
w' & = \frac{\alpha_1}{z}(w-1)^2+\frac{(1-A)}{2z}(w-1)+\frac{vw}{2z},\label{vw2}
\end{align}
\end{subequations}
where the derivatives are now with respect to $z$. Then, we clear $v$ from equation~\eqref{vw2}, derive the resulting equation and substitute $v$ and $v'$ in equation~\eqref{vw1}. After some long calculations we finally obtain one equation for $w$ given by
\begin{equation}
w''=\left(\frac{1}{2w}+\frac{1}{w-1}\right)(w')^2-\frac{w'}{z}+\frac{(w-1)^2}{z^2}\left(aw+\frac{b}{w}\right)+c\frac{w}{z}+d\frac{w(w+1)}{w-1},\label{PV}
\end{equation}
with the parameters
\begin{equation}
a=\frac{\alpha_1^2}{2},\quad b=-\frac{\alpha_3^2}{2},\quad c=\frac{\alpha_2-\alpha_4}{2},\quad d=-\frac{1}{8},\label{paraPV}
\end{equation}
which is the PV equation. In general $w\in\mathbb{C}$ and also the parameters $a,b,c,d\in\mathbb{C}$.

The corresponding spectrum contains four independent equidistant ladders starting from the extremal states \cite{CFNN04}:
\begin{subequations}
\begin{align}
\psi_{\mathcal{E}_1} \propto &\left[\frac{h}{2}\left(\frac{g'}{2g}-\frac{h'}{2h}-\frac{x}{2}-\frac{\alpha_1}{g}\right)-\alpha_1-\alpha_2-\frac{\alpha_3}{2}\right]\nonumber\\
&\times\exp\left[\int\left(\frac{g'}{2g}+\frac{g}{2}-\frac{\alpha_1}{g}\right)\text{d}x\right],\\
\psi_{\mathcal{E}_2} \propto & \left[\frac{h}{2}\left(\frac{g'}{2g}-\frac{h'}{2h}-\frac{x}{2}+\frac{\alpha_1}{g}\right)-\alpha_2-\frac{\alpha_3}{2}\right]\nonumber\\
& \times\exp\left[\int\left(\frac{g'}{2g}+\frac{g}{2}+\frac{\alpha_1}{g}\right)\text{d}x\right],\\
\psi_{\mathcal{E}_3} \propto & \exp\left[\int\left(\frac{h'}{2h}+\frac{h}{2}-\frac{\alpha_3}{h}\right)\text{d}x\right],\label{extRO3}\\
\psi_{\mathcal{E}_4} \propto & \exp\left[\int\left(\frac{h'}{2h}+\frac{h}{2}+\frac{\alpha_3}{h}\right)\text{d}x\right],\label{extRO4}
\end{align}\label{extremalRO}
\end{subequations}
\hspace{-1mm}where
\begin{equation}
h(x)=-x-g(x).\label{hdeg}
\end{equation}

The number operator $N_4(H)$ for this system will be a fourth-order polynomial in $H$. From the definitions in equations (\ref{elesRO}--\ref{cloconRO}) we can obtain the energies of the extremal states in terms of the factorization energies as 
$\mathcal{E}_i=\epsilon_i+1, \ i=1,2,3,4$.

Therefore, if we have a solution $w$ of the PV equation~\eqref{PV}, we obtain a system characterized by a third-order PHA.

\begin{table}
\begin{center}
\begin{tabular}{ccl}
\hline
PHA ($m$) & Ladder operators & System\\
\hline
0th-order&1st-order&Harmonic oscillator (HO)\\
1st-order&2nd-order&Radial oscillator (RO)\\
2nd-order&3rd-order&Connected with PIV\\
3rd-order&4th-order&Connected with PV\\
\hline
\end{tabular}
\vspace{-3mm}
\caption{\small{The first four PHA and their associated systems.}} \label{tablepha}
\end{center}
\end{table}

\section{SUSY QM, harmonic oscillator and PIV equation}
\label{cappain}
In section \ref{secondPHA} we saw that, in order to have a system described by second-order PHA, we needed to find solutions of PIV equation. Nevertheless, in this work we will use this connection in the reverse direction, i.e., first we look for systems which are certainly described by second-order PHA and then we develop a method to find solutions of the PIV equation.
\subsection{First-order SUSY partners of the harmonic oscillator}\label{1susyp4}
For $k=1$ we realized that the ladder operators $L_1^\pm$ are of third order. This means that the first-order SUSY transformation applied to the harmonic oscillator could provide solutions to the PIV equation. To find them, we need to identify first the extremal states, which are annihilated by $L_1^-$ and at the same time are eigenstates of $H_1$. From the corresponding spectrum, it is clear that the transformed ground state of $H_0$ and the eigenstate of $H_1$ associated with $\epsilon_1$ are two physical extremal states associated with our system. Since the other root of $N_3(H_1)$ is $\epsilon_1 + 1 \not\in{\rm Sp}(H_1)$, then  the corresponding extremal state will be non-physical, which can be simply constructed from the non-physical seed solution used to implement the transformation as $A_1^+ a^{+} u_1$. Due to this, the three extremal states for our system and their corresponding factorization energies (see equation~\eqref{q3}) become
\begin{subequations}
\begin{alignat}{5}
\psi_{{\cal E}_1} & \propto A_1^+ e^{-x^2/2}, &\quad \psi_{{\cal E}_2} & \propto A_1^+ a^{+} u_1, &\quad \psi_{{\cal E}_3} & \propto \frac{1}{u_1},\\
 {\cal E}_1 &= \frac{1}{2}, &\quad {\cal E}_2 &= \epsilon_1 + 1, &\quad {\cal E}_3 &= \epsilon_1.
\end{alignat}
\end{subequations}

\begin{figure}
\begin{center}
\includegraphics[scale=0.335]{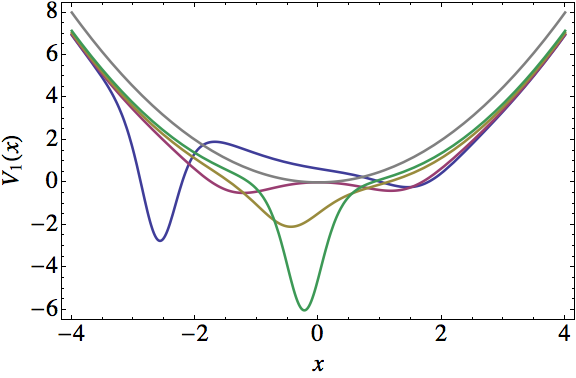} \hskip0.4cm
\includegraphics[scale=0.35]{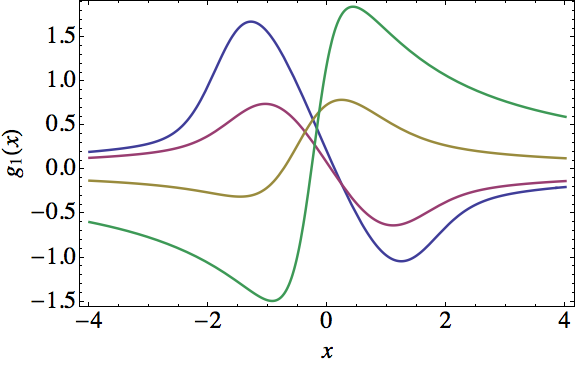}
\end{center}
\vspace{-5mm}
\caption{\small{First-order SUSY partner potentials $V_1(x)$ (left) of the harmonic oscillator and the PIV solutions $g_1(x,\epsilon_1)$ (right) for: $\epsilon_1 = 0.25$, $\nu_1=0.99$ (blue); $\epsilon_1=0$, $\nu_1=0.1$ (magenta); $\epsilon_1=-1$, $\nu_1=0.5$ (yellow); and $\epsilon_1=-4$, $\nu_1=0.5$ (green).}}\label{PIV1}
\end{figure}

The first-order SUSY partner potential $V_1(x)$ and the non-singular solution of the PIV equation are
\begin{subequations}
\begin{align}
V_1(x) &= \frac{x^2}2 - \{\ln [u_1(x)]\}', \\
 g_1(x,\epsilon_1) &=  - x - \{\ln [\psi_{{\cal E}_3}(x)]\}' = - x  + \{\ln[u_1(x)]\}',\label{solg1a}
\end{align}
\end{subequations}
where we label the PIV solution with an index characterizing the order of the transformation employed and we explicitly indicate the dependence on the factorization energy.  Note that two additional solutions of the PIV equation can be obtained by cyclic permutations of the indices $(1,2,3)$. However, they will have singularities at some points and thus we drop them in this approach. An illustration of the first-order SUSY partner potentials $V_1(x)$ of the harmonic oscillator as well as its corresponding PIV solutions $g_1(x,\epsilon_1)$ are shown in figure~\ref{PIV1}.

\subsection{Reduction theorem and third-order ladder operators}\label{sectma}
We just saw that the 1-SUSY partners of the harmonic oscillator are ruled by second-order PHA, but we can ask if there are any other systems with this kind of algebra? In this section we present a {\it reduction theorem} in which it is shown that special families of $k$th-order SUSY partners of the harmonic oscillator, normally ruled by $2k$th-order algebras, can also have second-order ones. The proof of this theorem can be found in \cite{BF11a}.\\

\noindent {\bf Theorem 1.} Suppose that the $k$th-order SUSY partner $H_k$ of the harmonic oscillator Hamiltonian $H_0$ is generated by $k$ Schr\"odinger seed solutions, which are connected by the standard annihilation operator in the way:
\begin{equation}
u_j = (a^{-})^{j-1} u_1, \quad \epsilon_j = \epsilon_1 - (j-1), \quad j=1,\dots,k, \label{restr}
\end{equation}
where $u_1(x)$ is a nodeless Schr\"odinger solution given by equation~\eqref{hyper} for $\epsilon_1 < 1/2$ and $\vert \nu_1 \vert < 1$.
\smallskip
Therefore, the natural ladder operator $L_k^+ = B_k^{+} a^{+} B_k^{-}$ of $H_k$, which is of $(2k+1)$th-order, is factorized in the form
\begin{equation}
L_k^+ = P_{k-1}(H_k) l_k^+,\label{hipo}
\end{equation}
where $P_{k-1}(H_k) = (H_k - \epsilon_1)\dots(H_k - \epsilon_{k-1})$ is a polynomial of $(k-1)$th-order in $H_k$, $l_k^+$ is a third-order differential ladder operator such that
\begin{equation}
[H_k,l_k^+] = l_k^+, \label{conmHl}
\end{equation}
and
\begin{equation} \label{annumk3}
 l_k^+ l_k^- = (H_k - \epsilon_k)\left(H_k - \frac{1}{2} \right)(H_k - \epsilon_1 - 1).
\end{equation}

\subsection{Solutions of the PIV equation departing from $H_k$}\label{realsols}
We discussed in section~\ref{pha} that there are some PHA that are \emph{reducible}, i.e., they fulfill  $\mathcal{L}_m^{+}={\cal P}(H)a^{+}$. In the case addressed by Theorem 1 we have a similar situation, i.e., when the SUSY transformation fulfills the requirements given there, the algebra generators become factorized as in equation~\eqref{hipo}. This means that the $2k$th-order PHA, obtained through a $k$th-order SUSY transformation as specified in the theorem, with $\epsilon_j=\epsilon_1-(j-1),\ j=1,\dots ,k$, will be \emph{reduced} to a second-order PHA with third-order ladder operators.

This implies that when we reduce the higher-order algebras the possibility is open of generating new solutions of the PIV equation. This happens indeed: first we obtained solution families already given in the literature \cite{BF11a}, then we worked to expand the solution space in the parameters $a,b$. We have generated real solutions associated with real parameters \cite{BF11a}, then complex solutions associated with real parameters \cite{BF12,BF13a} and, finally, complex solutions associated with complex parameters \cite{Ber12}. In this section we will show the method used to obtain these solutions and next we classify them into solution hierarchies \cite{BF11a,BF13a}.

\begin{figure}[H]
\begin{center}
\includegraphics[scale=0.335]{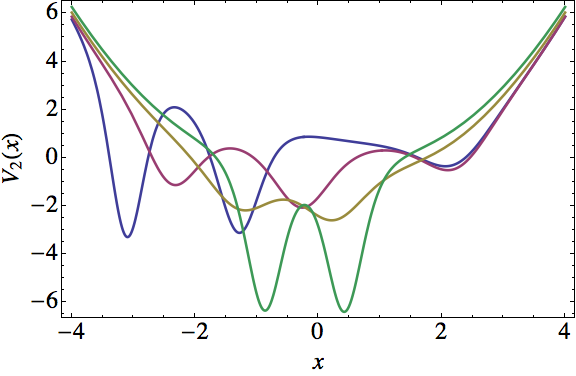} \hskip0.4cm
\includegraphics[scale=0.35]{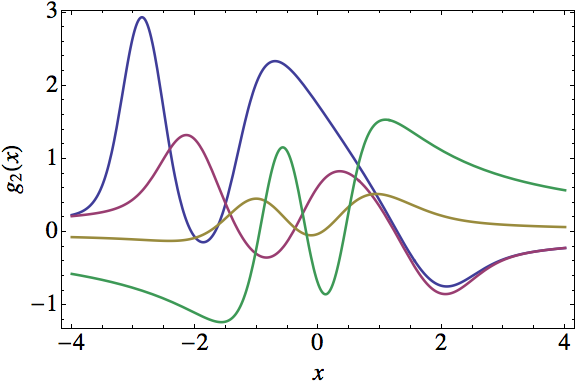}
\end{center}
\vspace{-5mm}
\caption{\small{Second-order SUSY partner potentials $V_2(x)$ (left) of the harmonic oscillator and the corresponding PIV solutions $g_2(x,\epsilon_1)$ (right) for $\epsilon_1 = 0.25,\nu_1=0.99$ (blue), and $\epsilon_1=\{0.25 \text{ (magenta)}$,$-0.75\text{ (yellow)}$,$-2.75\text{ (green)}\}$ with $\nu_1 = 0.5$.}}\label{PIV2}
\end{figure}

\begin{figure}[H]
\begin{center}
\includegraphics[scale=0.335]{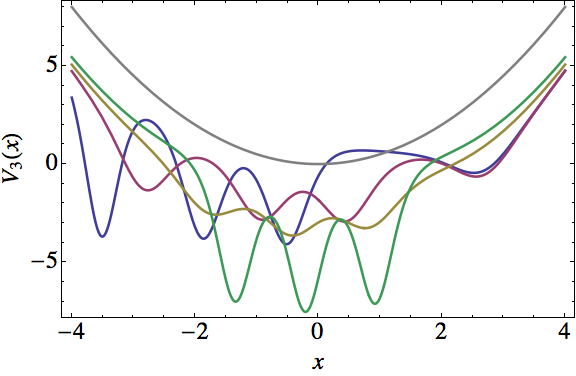} \hskip0.4cm
\includegraphics[scale=0.35]{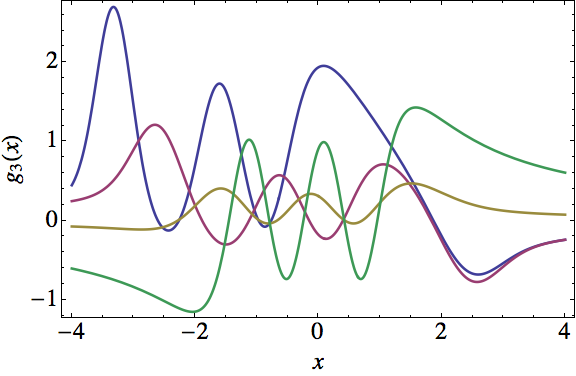}
\end{center}
\vspace{-5mm}
\caption{\small{Third-order SUSY partner potentials $V_3(x)$ (left) of the harmonic oscillator and the corresponding PIV solutions $g_3(x,\epsilon_1)$ (right) for $\epsilon_1 = 0.25,\nu_1=0.99$ (blue), and $\epsilon_1=\{0.25 \text{ (magenta)}$,$-0.75\text{ (yellow)}$,$-2.75\text{ (green)}\}$ with $\nu_1 = 0.5$.}}\label{PIV3}
\end{figure}

In order to get the PIV solutions we need to identify the extremal states of our system. Since the roots of the polynomial of equation~\eqref{annumk3} are now $E_0,\, \epsilon_1 + 1,$ and $\epsilon_k = \epsilon_1 - (k-1)$, then the spectrum of $H_k$ consists of two physical ladders: an infinite one departing from $E_0$ and a finite one starting from $\epsilon_k$ and ending at $\epsilon_1$. Thus, there are two physical extremal states corresponding to the mapped ground state of $H_0$ with eigenvalue $E_0$ and the eigenstate of $H_k$ associated with $\epsilon_k $. The other extremal state (which corresponds to $\epsilon_1 +1 \not\in{\rm Sp}(H_k)$) is non-physical. Finally, the three extremal states are
\begin{subequations}
\begin{alignat}{5}
\psi_{{\cal E}_1} & \propto B_k^+ e^{-x^2/2}, \quad & \psi_{{\cal E}_2} & \propto B_k^+ a^{+} u_1,  \quad &
\psi_{{\cal E}_3} & \propto \frac{W(u_1,\dots,u_{k-1})}{W(u_1,\dots,u_k)}, \label{edo1}\\
 {\cal E}_1& = \frac{1}{2}, \quad &  \quad {\cal E}_2 & = \epsilon_1 + 1,\quad & {\cal E}_3 & = \epsilon_k = \epsilon_1 - (k - 1).
\end{alignat}\label{edo123}
\end{subequations}
\hspace{-1mm}The $k$th-order SUSY partner potential of the harmonic oscillator and the corresponding non-singular solution of the PIV equation become
\begin{subequations}
\begin{align} \label{parafig3}
V_k(x) &= \frac{x^2}2 - \{\ln [W(u_1,\dots,u_k)]\}'' , \quad k\geq 2,\\
g_k(x,\epsilon_1) &= - x - \{\ln[\psi_{{\cal E}_3}(x)]\}' =  - x - \left\{\ln \left[\frac{W(u_1,\dots,u_{k-1})}{W(u_1,\dots,u_{k})}\right]\right\}'.\label{solg}
\end{align}\label{parafig3ysolg}
\end{subequations}
\hspace{-1.5mm}We have illustrated the $k$th-order SUSY partner potentials $V_k(x)$ of the harmonic oscillator and the corresponding PIV solutions $g_k(x,\epsilon_1)$ in figure \ref{PIV2} for $k=2$ and in figure~\ref{PIV3} for $k=3$.

Using this algorithm we are able to find solutions to the PIV equation with specific parameters $a,b$, i.e., not for any combination of them. Actually, we can express $a,b$ in terms of the two parameters of the transformation $\epsilon_1,k$. However, as $a,b,\epsilon_1\in\mathbb{R}$ but $k\in\mathbb{Z}^{+}$, we cannot expect to cover all the parameter space $a,b$. Let us note that the same set of real solutions to the PIV equation can be obtained through inverse scattering techniques \cite{AC92} (compare the solutions of \cite{BCH95} with those of \cite{BF11a}). Indeed, we have
\begin{equation}
a=-\epsilon_1+2k-\frac{3}{2},\qquad b=-2\left(\epsilon_1+\frac{1}{2}\right)^2.
\end{equation}

Let us intend to overcome now the restriction $\epsilon_1<E_0$, although if we use the formalism as in \cite{BF11a}, we would only obtain singular SUSY transformations. In order to avoid this, we will instead employ complex SUSY transformations. The simplest way to implement them is to use a complex linear combination of the two standard linearly independent real solutions which, up to an unessential factor, leads to the following complex solutions depending on a complex constant $\lambda + i \kappa$ ($\lambda, \kappa \in \mathbb{R}$) \cite{AICD99}:
\begin{equation}
u(x;\epsilon ) = e^{-x^2/2}\left[ {}_1F_1\left(\frac{1-2\epsilon}{4},\frac12;x^2\right)
 + x(\lambda + i\kappa)\, {}_1F_1\left(\frac{3-2\epsilon}{4},\frac32;x^2\right)\right]. \label{u1}
\end{equation}
The known real results are obtained by taking $\kappa=0$ and expressing $\lambda$ as \cite{JR98} (with $\nu \in \mathbb{R}$):
\begin{equation}
\lambda= 2 \nu\frac{\Gamma(\frac{3 - 2\epsilon}{4})}{\Gamma(\frac{1-2\epsilon}{4})}. \label{nu}
\end{equation}

Hence, through this formalism we will obtain $V_k(x)$ and their corresponding $g_k(x;\epsilon_1)$, both of which will now be complex, using once again equations \eqref{parafig3ysolg}. In addition, the extremal states of $H_{k}$ and their corresponding energies are given by equations \eqref{edo123}. Recall that all $u_j$ satisfy equation~\eqref{restr} and $u_1$ corresponds to the general solution given in equation~\eqref{u1}. 

\begin{figure}
\begin{center}
\includegraphics[scale=0.33]{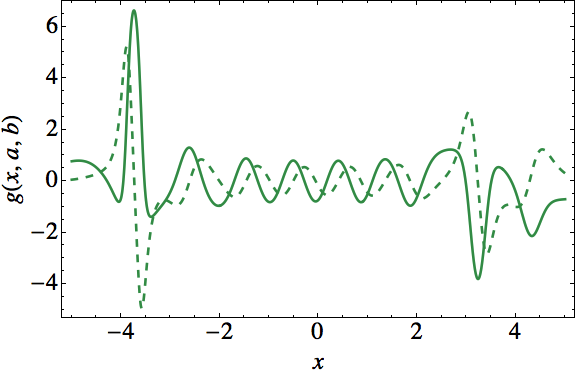}\hspace{5mm}
\includegraphics[scale=0.33]{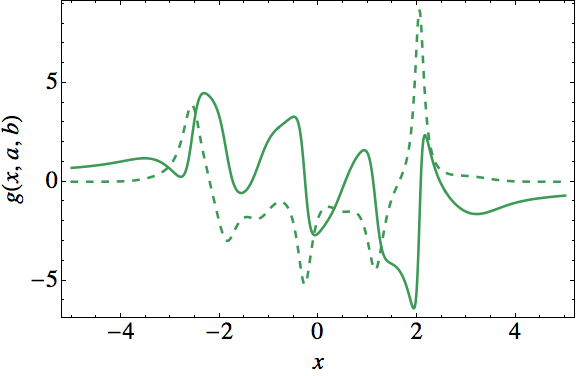}
\end{center}
\vspace{-5mm}
\caption{\small{Real (solid line) and imaginary (dashed line) parts of some complex solutions to PIV equation. The left plot corresponds to $a_{ii}=12$, $b_{ii}=-8$ ($k=2$, $\epsilon_1=7$, $\lambda=\kappa=1$) and the right one to $a_{iii}=-5$, $b_{iii}=-8$ ($k=1$, $\epsilon_1=5/2$, $\lambda=\kappa=1$).}} \label{gcomplex1}
\end{figure}

\begin{figure}
\begin{center}
\includegraphics[scale=0.33]{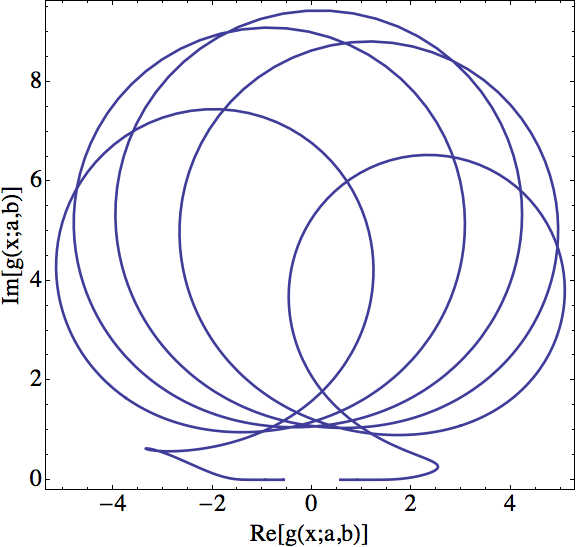}
\end{center}
\vspace{-5mm}
\caption{\small{Parametric plot of the real and imaginary parts of $g(x;a,b)$ for  $a_i=-9/2$, $b_i=-121/2$ ($k=1$, $\epsilon_1=5$, $\lambda=\kappa=2$) and $|x| \leq 10$. For larger values of $x$, the curve slowly approaches the origin from both sides.}} \label{complexpara1}
\end{figure}

Note that, in general, $\psi_{\mathcal{E}_j}\neq 0\ \forall\ x \in \mathbb{R}$ which implies that, by making cyclic permutations of the indices of the three energies $\mathcal{E}_j$ and the corresponding extremal states of equations~\eqref{edo123}, we expand the solution families to three different sets defined by
\begin{subequations}
\begin{alignat}{3}
a_{i}&=-\epsilon_1 + 2k -\frac{3}{2}, & \quad  b_{i} & =-2\left(\epsilon_1+\frac{1}{2}\right)^{2}, \label{ab1}\\
a_{ii}&= 2\epsilon_1 -k, & \quad b_{ii} & =-2k^2, \\
a_{iii}&=-\epsilon_1-k-\frac{3}{2}, & \quad b_{iii} & =-2\left(\epsilon_1 - k +\frac{1}{2}\right)^2,
\end{alignat}
\end{subequations}
where the new indices aim to distinguish between them for fixed values of $\epsilon_1$ and $k$. The corresponding PIV solutions $g(x;a,b)$ are not singular and their real and imaginary parts have a null asymptotic behaviour ($g\rightarrow 0$ as $|x|\rightarrow \infty$). This property appear clearly in the example of figure~\ref{gcomplex1}, and in the parametric plot of the real and imaginary parts of the $g(x;a,b)$ of figure~\ref{complexpara1}.

\begin{figure}
\begin{center}
\includegraphics[scale=0.33]{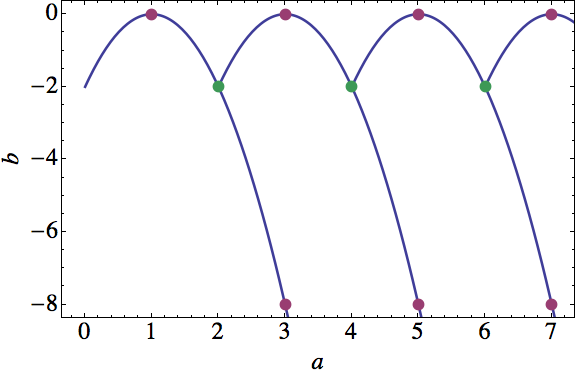}
\includegraphics[scale=0.33]{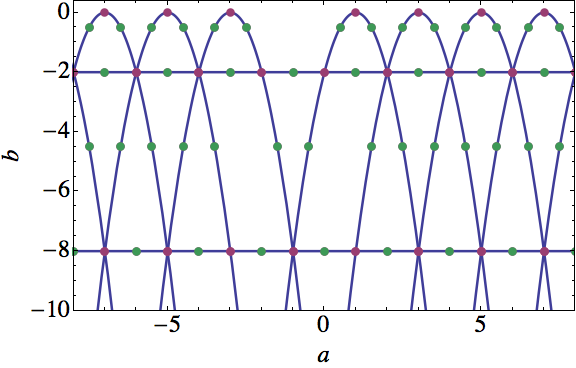}
\end{center}
\vspace{-5mm}
\caption{\small{Parameter space for real (left) and complex (right) solution hierarchies. The lines correspond to the confluent hypergeometric function, the dots to more specific hierarchies.}} \label{complexpara2}
\end{figure}
\subsection{PIV solution hierarchies}\label{realhie}
The solutions $g_k(x,\epsilon_1)$ of the PIV equation can be classified according to the explicit special functions of which they depend on \cite{BCH95,BF11a,Ber12,BF13a}. Our general formulas, given by equations~\eqref{solg}, are expressed in terms of the confluent hypergeometric function, although for particular values of the parameter $\epsilon_1$ they can be simplified to more specific special functions.

\begin{figure}
\includegraphics[scale=0.34]{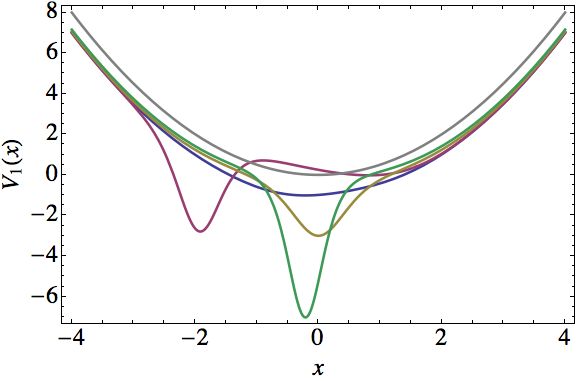}
\includegraphics[scale=0.34]{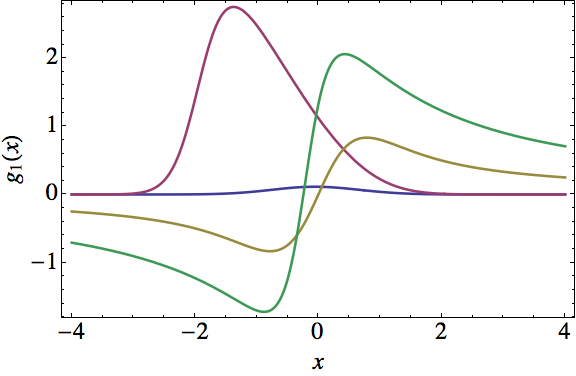}
\vspace{-5mm}
\caption{\small{First-order SUSY partner potentials $V_1(x)$ (left) of the harmonic oscillator and the PIV solutions $g_1(x,\epsilon_1)$ (right) belonging to the error function hierarchy for: $\epsilon_1 =-0.5,\nu_1=0.1$ (blue); $\epsilon_1 =-0.5,\nu_1=0.99$ (magenta); $\epsilon_1 =-1.5,\nu_1=0.001$ (yellow); and $\epsilon_1 =-3.5,\nu_1=0.5$ (green).}}\label{PIVerf}
\end{figure}

Let us remark that, in this work we are interested in non-singular SUSY partner potentials and their corresponding non-singular solutions of the PIV equation. We can obtain both real and complex non-singular solutions for certain parameters $a,b$ of the PIV equation. In figure~\ref{complexpara2} we show the parameter space where solutions can be found. For real solutions we have identified the following hierarchies, which lie on specific points of the parameter space.
\begin{itemize}
\item Confluent hypergeometric function hierarchy
\begin{align}\displaystyle
g_1(x,\epsilon_1)=&  \frac{2\nu_1\Gamma\left(\frac{3-2\epsilon_1}{4}\right) \left[(3-6x^2)\,{}_1F_1\left(\frac{3-2\epsilon_1}{4},\frac{3}{2};x^2\right)+x^2(3-2\epsilon_1)\,{}_1F_1\left(\frac{7-2\epsilon_1}{4},\frac{5}{2};x^2\right) \right]}
{3 \Gamma\left(\frac{1-2\epsilon_1}{4}\right)\,{}_1F_1\left(\frac{1-2\epsilon_1}{4},\frac{1}{2};x^2\right) + 6\nu_1 x \Gamma\left(\frac{3-2\epsilon_1}{4}\right)\,{}_1F_1(\frac{3-2\epsilon_1}{4},\frac{3}{2};x^2) }\nonumber\\
&+\frac{x\Gamma\left(\frac{1-2\epsilon_1}{4}\right)\left[ -2\,{}_1F_1\left(\frac{1-2\epsilon_1}{4},\frac{1}{2};x^2\right)+(1-2\epsilon_1)\,{}_1F_1\left(\frac{5-2\epsilon_1}{4},\frac{3}{2};x^2\right)	 \right]}
{\Gamma\left(\frac{1-2\epsilon_1}{4}\right)\,{}_1F_1\left(\frac{1-2\epsilon_1}{4},\frac{1}{2};x^2\right) + 2\nu_1 x \Gamma\left(\frac{3-2\epsilon_1}{4}\right)\,{}_1F_1(\frac{3-2\epsilon_1}{4},\frac{3}{2};x^2) }. \label{g1}
\end{align}
\item Error function hierarchy (some SUSY partner potentials and the PIV solutions corresponding to this hierarchy are shown in figure~\ref{PIVerf})
\begin{subequations}
\begin{align}
g_1(x,-5/2)&=\frac{4[\nu_1 + \varphi_{\nu_1}(x)]}{2\nu_1 x +(1+2x^2)\varphi_{\nu_1}(x)},\label{erf2}\\
g_2(x,-1/2)&=\frac{4\nu_1[\nu_1 + 6\varphi_{\nu_1}(x)]}{\varphi_{\nu_1}(x)[\varphi_{\nu_1}^2(x) -2\nu_1 x \varphi_{\nu_1}(x) -2\nu_1^2]}.
\end{align}\label{erfs}
\end{subequations}
\item Rational hierarchy
\begin{subequations}
\begin{align}
g_2(x,-9/2)&= -\frac{8 (3 x + 2 x^3)}{3 + 12 x^2 + 4 x^4} + \frac{32 (15 x^3 + 12 x^5 + 4 x^7)}{45 + 120 x^4 + 64 x^6 + 16 x^8},\\
g_3(x,-5/2)=& \frac{4 x (27 - 72 x^2 + 16 x^8)}{27 + 54 x^2 + 96 x^6 - 48 x^8 + 32 x^{10}}.\label{pol3}
\end{align}
\end{subequations}
\end{itemize}

\section{SUSY QM, radial oscillator and PV equation}
\label{5painleve}
Here we are going to follow the same procedure of section \ref{cappain}, in which we used the SUSY partners of the harmonic oscillator to generate solutions of the PIV equation, but now employing the radial oscillator SUSY partners to produce solutions of the PV equation.
\subsection{First-order SUSY partners of the radial oscillator}
If we calculate the first-order SUSY partners of the radial oscillator, we get a system naturally ruled by a third-order PHA. We will obtain now the solutions of the PV equation following \cite{CFNN04}. To do that, we need to identify the extremal states of $H_1$ and its associated energies. From the spectrum of the radial oscillator we have already two extremal states, one physical associated with $E_0=j/2+3/4$ and one non-physical for $-E_0+1$\footnote{In this section we switch to $j$ the angular momentum index since we will use $\ell$ to denote the reduced ladder operators for the radial oscillator Hamiltonian (see also \cite{Ber13}).}. The other two roots are added by the SUSY transformation, one for the new level at $\epsilon_1$ and the other at $\epsilon_1+1$ to have a finite ladder, namely,
\begin{subequations}
\begin{alignat}{3}
\psi_{{\cal E}_1} & \propto A_1^+b^+u, &\quad {\cal E}_1 & = \epsilon_1+1,\\
\psi_{{\cal E}_2} & \propto A_1^+ \left[x^{-j} \exp(-x^2/4)\right], & \quad  {\cal E}_2 & = -E_0+1,\\
\psi_{{\cal E}_3} & \propto u^{-1}, & \quad  {\cal E}_3 & = \epsilon_1,\\
\psi_{{\cal E}_4} & \propto A_1^+ \left[x^{j+1} \exp(-x^2/4)\right], & \quad  {\cal E}_4 & = E_0,
\end{alignat}\label{extremalp5}
\end{subequations}
\hspace{-1.5mm}$A_1$ and $b^+$ being the first-order intertwining and ladder operators for the radial oscillator respectively.

For this system we are able to connect with PV equation with specific parameters $a,b,c,d\in\mathbb{C}$. From equations \eqref{paraPV} and \eqref{extremalp5} we obtain $a,b,c,d$ in terms of one parameter of the original system $E_0$ and one of the SUSY transformation $\epsilon_1$ as
\begin{equation}
a=\frac{(E_0+\epsilon_1)^2}{2},\quad b=-\frac{(E_0-\epsilon_1)^2}{2},\quad c=\frac{1-2E_0}{2},\quad d=-\frac{1}{8}.
\end{equation}
Since $E_0=E_0(j)=j/2+3/4$, then we can express the parametrization as
\begin{equation}
a=\frac{(2j+4\epsilon_1+3)^2}{32},\quad b=-\frac{(2j-4\epsilon_1+3)^2}{32},\quad c=-\frac{2j+1}{4},\quad d=-\frac{1}{8}.
\end{equation}
In general, the four parameters $a,b,c,d$ are written in terms of $j\in\mathbb{R}^+$ and $\epsilon_1\in\mathbb{C}$, although in this section we study the case where both are real, i.e., $a,b,c,d,\epsilon_1\in\mathbb{R}$. We must remark that in the physical studies of the radial oscillator systems usually $j\in\mathbb{Z}^+$, as it is the angular momentum index. However, here we will consider $j\in\mathbb{R}^+$ because we only use it as an auxiliary system to obtain solutions to PV equation.

\begin{figure}\centering
\includegraphics[scale=0.34]{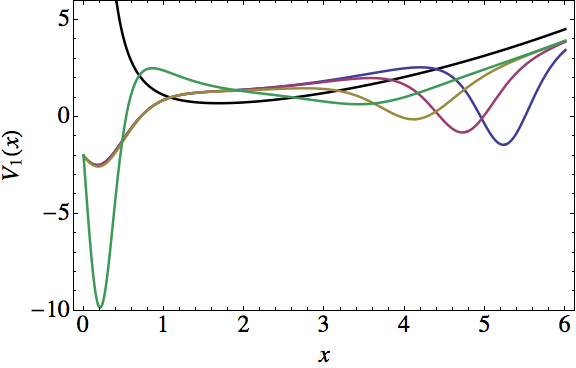}
\includegraphics[scale=0.34]{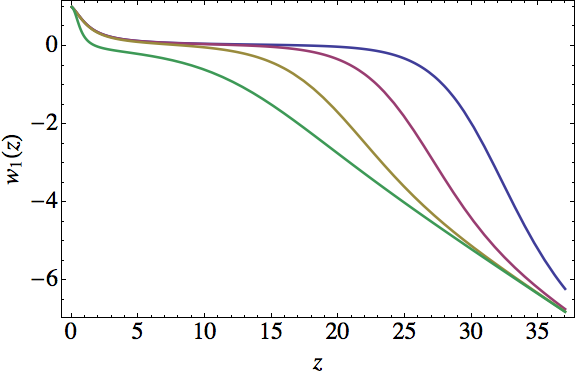}
\caption{\small{SUSY partner potential $V_1(x)$ of the radial oscillator (black) (left) and the PV solutions $w_1(z)$ (right) for $j=1$, $\epsilon_1=1$, and $\nu_1=\{$0.905 (blue), 0.913 (magenta), 1 (yellow), 10 (green)$\}$.}}\label{pv1}
\end{figure}

If we restrict ourselves to non-singular real solutions of the PV equation with real parameters we also have the restriction $\epsilon_1\leq E_0=j/2+3/4$. Moreover, for each value of $\epsilon_1$ we have a one-parameter family of solutions, labelled by the parameter $\nu_1$ from equation~\eqref{solRO} under the restrictions of equation~\eqref{condRO}. Then from 1-SUSY we can obtain the following partner potential and the function $g(x)$ related with the solution $w(z)$ of the PV equation,
\begin{equation}
V_1(x)=\frac{x^2}{8}+\frac{j(j+1)}{2x^2}-[\ln u(x)]'',\qquad
g_1(x)=-\frac{x}{2}-\frac{j+1}{x}+[\ln u(x)]',\end{equation}
where we have added an index to indicate the order of the SUSY transformation. Since $g_1(x)$ is connected with the solution $w_1(z)$ of PV equation through
\begin{equation}
w_1(z)=1+\frac{z^{1/2}}{g_1(z^{1/2})},\label{p5k}
\end{equation}
then
\begin{equation}
w_1(z)=1+\frac{2zu(z^{1/2})}{2z^{1/2}u'(z^{1/2})-(z+2j+1)u(z^{1/2})}.
\end{equation}
An illustration of the first-order SUSY partner potentials of the radial oscillator $V_1(x)$ and the corresponding solutions $w_1(z)$ of the PV equation are shown in figure~\ref{pv1}.

\subsection{Reduction theorem and fourth-order ladder operators}\label{sectmaRO}
Now we will show that some odd-order PHA associated with the SUSY partners of the radial oscillator can be reduced to third-order ones, which are generated by fourth-order ladder operators. To accomplish that, in this section we will present a new {\it reduction theorem}, through which we will identify the special higher-order SUSY partners of the radial oscillator, normally ruled by a $(2k+1)$th-order PHA but having associated also a third-order one. Recall that in this section we are using $j$ as the angular momentum index in order to free $\ell$, to be employed as the reduced ladder operator. We also stop writing explicitly the dependence of the radial oscillator Hamiltonian $H_0$, its eigenvalues $E_n$, and its ladder operators $b^\pm$ on the angular momentum index. The proof of this theorem can be found in \cite{Ber13}.\\

\noindent {\bf Theorem 2.} Let $H_k$ be the $k$th-order SUSY partner of the radial oscillator Hamiltonian $H_0$, generated by $k$ Schr\"odinger seed solutions. Suppose that these solutions $u_i$ are connected by the annihilation operator of the radial oscillator $b^-$ as
\begin{equation}
u_i = (b^{-})^{i-1} u_1, \quad \epsilon_i = \epsilon_1 - (i-1), \quad i=1,\dots,k, \label{restrRO}
\end{equation}
where $u_1(x)$ is a nodeless Schr\"odinger solution given by \eqref{solRO} for $\epsilon_1 < E_0=j/2+3/4$ and
\begin{equation}
\nu_1\geq -\frac{\Gamma\left(\frac{1-2j}{2}\right)}{\Gamma\left(\frac{1-2j-4\epsilon_1}{4}\right)}.
\end{equation}
Therefore, the natural $(2k+2)$th-order ladder operator $L_k^+ = B_k^{+} b^{+} B_k^{-}$ of $H_k$ turn out to be factorized in the form
\begin{equation}
L_k^+ = P_{k-1}(H_k) \ell_k^+,\label{hipoRO}
\end{equation}
where $P_{k-1}(H_k) = (H_k - \epsilon_1)\dots(H_k - \epsilon_{k-1})$ is a polynomial of $(k-1)$th-order in $H_k$ and $\ell_k^+$ is a fourth-order differential ladder operator,
\begin{equation}
[H_k,\ell_k^+] = \ell_k^+, \label{conmHlRO}
\end{equation}
such that
\begin{equation} \label{annumk3RO}
\ell_k^+ \ell_k^- =\left(H_k -E_0 \right)\left(H_k + E_0 -1 \right)(H_k - \epsilon_k)(H_k - \epsilon_1-1).
\end{equation}

\subsection{Solutions of PV equation departing from $H_k$}

Through Theorem 2 we are able to reduce the ($2k+1$)th-order PHA, induced by the natural ladder operators for the SUSY partners of the radial oscillator, to third-order PHA generated by fourth-order ladder operators. Basically, the $k$ transformation functions have to be connected by the annihilation operator $b^-$ and, therefore, their energies will be given by $\epsilon_i=\epsilon_1-(i-1)$. This means that, to build $H_k$, we have to create one equidistant ladder with $k$ steps, one for each first-order SUSY transformation. There is also the restriction on the free factorization energy that $\epsilon_1<E_0$. Thus, the possibility is open of obtaining new solutions to the PV equation, similar to the case of second-order PHA and PIV equation.

First we need to identify the extremal states of our system. The roots of the polynomial in \eqref{annumk3RO} are
$E_0, -E_0 +1, \epsilon_k, \epsilon_1+1$, two of them associated to physical extremal states ($E_0$, $\epsilon_k$), a non-physical one coming from the radial oscillator ($-E_0+1$), and another non-physical that will make the new ladder to be finite ($\epsilon_1+1$). The four extremal states are thus
\begin{subequations}
\begin{alignat}{3}
\psi_{{\cal E}_1} & \propto B_k^+b^+u_1, & \quad {\cal E}_1 & = \epsilon_1+1,\\
\psi_{{\cal E}_2} & \propto B_k^+ \left[x^{-j} \exp(-x^2/4)\right], & \quad  {\cal E}_2 & = -E_0+1,\\
\psi_{{\cal E}_3} & \propto \frac{W(u_1,\dots ,u_{k-1})}{W(u_1,\dots ,u_k)}, & \quad  {\cal E}_3 & = \epsilon_k,\\
\psi_{{\cal E}_4} & \propto B_k^+ \left[x^{j+1} \exp(-x^2/4)\right], & \quad  {\cal E}_4 & = E_0.
\end{alignat}\label{extremalp5k}
\end{subequations}
\hspace{-1mm}To simplify calculations we are going to use that
\begin{equation}
\psi_{{\cal E}_4} \propto B_k^+ \left[x^{j+1} \exp(-x^2/4)\right]\propto \frac{W(u_1,\dots , u_k,x^{j+1}\exp(-x^2/4))}{W(u_1,\dots , u_k)}.
\end{equation}
Then, from equations \eqref{extremalRO} we obtain the auxiliary function $h(x)$ defined in \eqref{hdeg},
\begin{equation}
h(x)=\left\{\ln\left[W(\psi_{\mathcal{E}_3},\psi_{\mathcal{E}_4})\right]\right\}',
\end{equation}
and for $g(x)$ it turns out that
\begin{equation}
g(x)=-x-h(x)=-x-\left\{\ln\left[W(\psi_{\mathcal{E}_3},\psi_{\mathcal{E}_4})\right]\right\}'. \label{g34}
\end{equation}

\begin{figure}\centering
\includegraphics[scale=0.34]{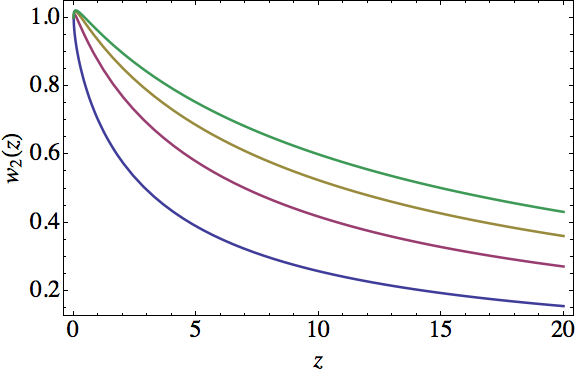}
\includegraphics[scale=0.34]{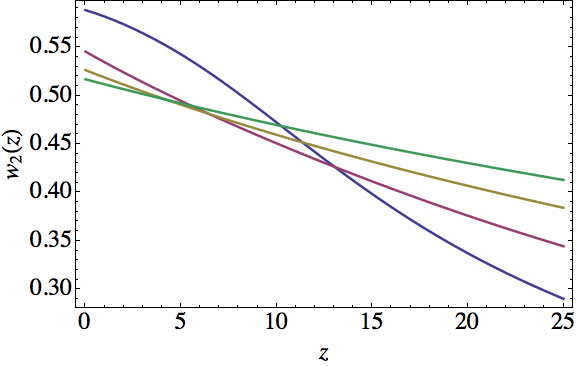}
\caption{\small{PV solutions $w_2$ generated by the second-order SUSY QM. The left plot is for the parameters $j=0$, $\nu_1=0$, and $\epsilon_1=\{1/4$ (blue), $-3/4$ (magenta), $-7/4$ (yellow), $-11/4$ (green)$\}$. The right plot is for $\epsilon_1=0$, $\nu_1=0$, and $j=\{$1 (blue), 3 (magenta), 6 (yellow), 10 (green)$\}$.}}\label{solskp5}
\end{figure}

Therefore, the $k$th-order SUSY partner potential $V_k(x)$ of the radial oscillator and its corresponding function $g_k(x)$ are
\begin{subequations}
\begin{align}
V_k(x)&=\frac{x^2}{8}+\frac{j(j+1)}{2x^2}-[\ln W(u_1,\dots , u_k)]'',\\
g_k(x)&=-x-\frac{2(E_0+\epsilon_1-k)W(u_1,\dots ,u_{k-1})W(u_1,\dots ,u_{k},x^{j+1}\exp(-x^2/4))}{W\left(W(u_1,\dots ,u_{k-1}), W(u_1,\dots ,u_{k},x^{j+1}\exp(-x^2/4))\right)}.\label{VgkROb}
\end{align}\label{VgkRO}
\end{subequations}
\hspace{-1mm}Recall that $g_k(x)$ is directly related with the function $w_k(z)$ as in equation \eqref{p5k},
which is a solution to PV equation with parameters given by
\begin{equation}
a=\frac{(E_0+\epsilon_1)^2}{2},\quad b=-\frac{(E_0-\epsilon_1+k-1)^2}{2},\quad c=\frac{k-2E_0}{2},\quad d=-\frac{1}{8}.
\end{equation}
In figure~\ref{solskp5} we show some PV solutions $w_2(z)$, obtained through the second-order SUSY transformation.

Note that we can use Theorem 2 even with complex transformation functions. The simplest way to implement them is to use a complex linear combination of two standard linearly independent real solutions with a complex constant $\lambda+i\kappa$, as
\begin{align}
u(x,\epsilon)=\, & x^{-j}\text{e}^{-x^2/4}\left[{}_1F_1\left(\frac{1-2j-4\epsilon}{4},\frac{1-2j}{2};\frac{x^2}{2}\right) \right.\nonumber\\
&+\left.(\lambda+i\kappa)\left(\frac{x^2}{2}\right)^{j+1/2}{}_1F_1\left(\frac{3+2j-4\epsilon}{4},\frac{3+2j}{2};\frac{x^2}{2}\right)\right].\label{solRO2}
\end{align}
The results obtained for the real solutions of equation~\eqref{solRO} are accomplished if we take
\begin{equation}
\lambda=\nu \frac{\Gamma\left(\frac{3+2j-4\epsilon}{4}\right)}{\Gamma\left(\frac{3+2j}{2}\right)},\quad \kappa=0.
\end{equation}

Comparing with the case when we were only looking for real solutions, now we have two restrictions that can be surpassed. The first of them is the inequality $\epsilon_1<E_0$, and the second one is that we had to choose our extremal states in the order of equation \eqref{extremalp5k}. Now we can perform permutations on the indices of the extremal states and we still do not obtain singularities, because in general $\psi_{\mathcal{E}_i}\neq 0$ in the complex plane.

In \eqref{paraPV} we have expressed the four parameters of the PV equation in terms of the four extremal state energies but we also have symmetry in the exchanges $\mathcal{E}_1 \leftrightarrow \mathcal{E}_2$ and $\mathcal{E}_3 \leftrightarrow \mathcal{E}_4$. Thus from the $4!=24$ possible permutations of the four indices we just have six different solutions to the PV equation. Next we show the parameters of the six solution families in terms of $\epsilon_1$, $j$, and $k$. We have added also an index to distinguish them
\begin{subequations}
\begin{alignat}{5}
a_1 & =\frac{(2j+4\epsilon_1+3)^2}{32}, & \ b_1 & =-\frac{(2j-4\epsilon_1+4k-1)^2}{32}, & \  c_1 & =\frac{-2j+2k-3}{4},\\
a_2 & =\frac{(2j+4\epsilon_1-4k+3)^2}{32}, & \ b_2 & =-\frac{(2j-4\epsilon_1-1)^2}{32}, & \  c_2 & =-\frac{2j+2k+1}{4},\\
a_3 & =\frac{(2j-4\epsilon_1+4k-1)^2}{32}, & \ b_3 & =-\frac{(2j+4\epsilon_1+3)^2}{32}, & \  c_3 & =\frac{2j-2k-1}{4},\\
a_4 & =\frac{(2j-4\epsilon_1-1)^2}{32}, & \ b_4 & =-\frac{(2j+4\epsilon_1- 4k+3)^2}{32}, & \  c_4 & =\frac{2j+2k+1}{4},\\
a_5 & =\frac{k^2}{2}, & \ b_5 & =-\frac{(2j+1)^2}{8}, & \  c_5 & =\frac{2\epsilon_1-k}{2},\\
a_6 & =\frac{(2j+1)^2}{8}, & \ b_6 & =-\frac{k^2}{2}, & \  c_6 & =-\frac{2\epsilon_1+k-1}{2}.
\end{alignat}\label{4paraRO} 
\end{subequations}

\begin{figure}\centering
\includegraphics[scale=0.34]{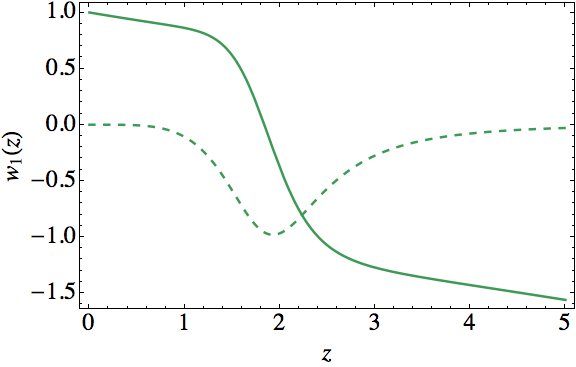}
\includegraphics[scale=0.34]{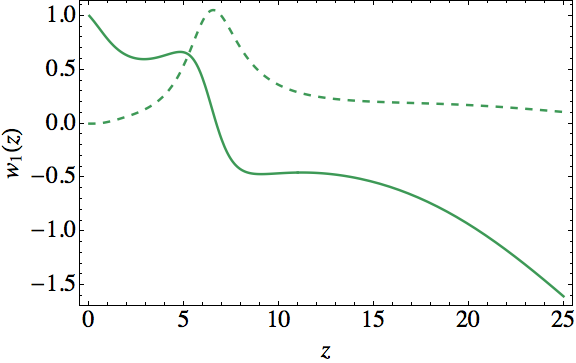}
\caption{\small{Real (solid) and imaginary (dashed) parts of the solution $w_1(z)$ to PV for $j=3$, $\epsilon_1=0$, and $\nu_1=100i$ (left) and $j=2$, $\epsilon_1=2$, and $\nu_1=i$ (right).}}\label{comp5}
\end{figure}

Then, the PV solutions are calculated from equation~\eqref{g34}. In figure~\ref{comp5} we show two complex solutions to the PV equation with real parameters $a,b,c,d$.

\begin{figure}\centering
\includegraphics[scale=0.34]{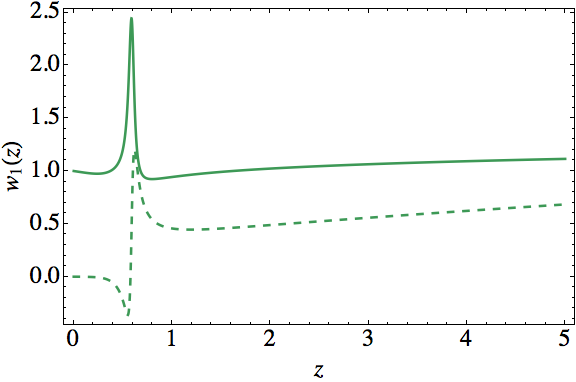}
\includegraphics[scale=0.34]{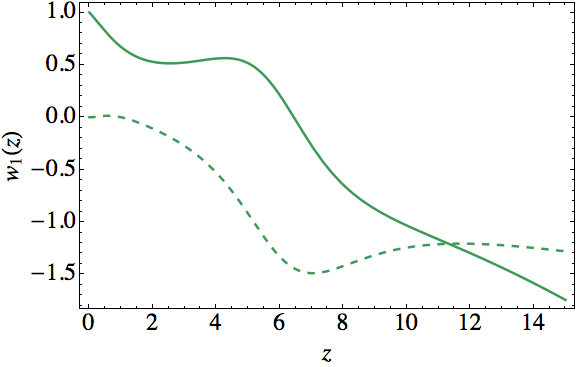}
\caption{\small{Real (solid) and imaginary (dashed) parts of the solution $w_1(z)$ to PV for $j=3$, $\epsilon_1=1+11i$, and $\nu_1=100i$; and $j=1$, $\epsilon_1=1-0.6i$, and $\nu_1=1-i$.}}\label{comp52}
\end{figure}

We can also obtain complex PV solutions simply by allowing the factorization {\it energy} in equation~\eqref{solRO2} to be complex. Then, these solutions will also be complex, as well as the parameters $a,b,c$ of the PV equation, as they depend on $\epsilon_1$. For example, in figure~\ref{comp52} we show two complex PV solutions but now associated with the following complex parameters:
\vspace{-2mm}
\begin{subequations}
\begin{alignat}{5}
a&=-\frac{115}{4}+i\frac{429}{16}, & \quad b& = \frac{1911}{32}+i\frac{55}{4}, & \quad c &= \frac{49}{4},\\
a&=-\frac{1881}{800}-i\frac{27}{20}, & \quad b& = \frac{119}{800}-i\frac{3}{20}, & \quad c &= -\frac{3}{4}.
\end{alignat}
\end{subequations}
\vspace{-7mm}
\subsection{PV solution hierarchies}
The solutions $w(z)$ that we have found for the PV equation are expressed in terms of the $g(x)$ in equation~\eqref{p5k}, and therefore in terms of the eigenfunctions $u_i$ of the radial oscillator (see e.g. equation~\eqref{VgkROb}). Recall that all $u_i$ are only determined by $\epsilon_1$ and $\nu_1$, due to the reduction theorem. The Painlev\'e equations themselves define new special functions, {\it the Painlev\'e trascendents}, which are defined as the general solutions of the corresponding equations. Nevertheless, for some special values of the parameters, they can be expressed in terms of known special functions. We can classify the solutions $w(z)$ of the PV equation into solution hierarchies, and some examples are the following.
\vspace{-2mm}
\begin{itemize}
\item Laguerre polynomials
\vspace{-3mm}\begin{equation}
w_1(z)= \ 1-z^{-1/2},\qquad
w_1(z)= \  1-\frac{z^{3/2}L_1^{(\alpha)}(z^2/2)}{2L_1^{(\alpha)}(z^2/2)-2\alpha-1},
\end{equation}
where $\alpha = -(2j + 1)/2$.
\item Hermite polynomials
\vspace{-3mm}\begin{subequations}\begin{align}
w_1(z)&=1-\frac{z^{3/2}H_{2n}(z)}{(z^2+1)H_{2n}(z)-4nzH_{2n-1}(z)},\label{w1hermitea}\\
w_1(z)&=1+\frac{z^{1/2}H_{2n}(z)}{4nH_{2n-1}(z)-zH_{2n}(z)}\label{w1hermiteb}.
\end{align}\label{w1hermite}\end{subequations}\vspace{-4mm}
\vspace{-3mm}\item Exponential function
\vspace{-3mm}\begin{subequations}
\begin{align}
w_1(z)&=1-\frac{z^{3/2}}{2}+\frac{z^{7/2}}{2z^{2}+4-4\exp(z^2/2)}.
\end{align}\label{w1expsol}
\end{subequations}
\vspace{-3mm}\item Modified Bessel functions
\vspace{-3mm}\begin{subequations}
\begin{align}
w_1(z)&=1+\frac{2I_\nu (z^2/4)}{z^{1/2}[I_{\nu+1}(z^2/4)-I_{\nu}(z^2/4)]}.
\end{align}\label{w1bessel}
\end{subequations}\vspace{-3mm}
\vspace{-3mm}\item Weber or Parabolic cylinder functions
\vspace{-3mm}\begin{subequations}
\begin{align}
w_1(z)&=1-\frac{2z^{3/2}E_\nu(z)}{2(z^2+1)E_\nu(z)-zE_{\nu-1}(z)+zE_{\nu+1}(z)},\\
w_1(z)&=1-\frac{2z^{1/2}E_\nu(z)}{2(z^2+1)E_\nu(z)-zE_{\nu-1}(z)+zE_{\nu+1}(z)}.
\end{align}
\end{subequations}\vspace{-3mm}
\end{itemize}
\vspace{-8mm}
\section{Conclusions}
In section 2, we studied the general framework of SUSY QM, from first to $k$th-order transformations and specifically the cases of the harmonic and radial oscillators. In section 3 we introduced the six Painlev\'e equations. Then, in section 4, we defined the PHA and in section 5, we obtained the general systems described by these algebras from zeroth to third order. In this direction, we would like to continue this study to higher-order systems.

After that, in sections 6 and 7 we formulated two {\it reduction theorems} for the higher-order SUSY partners of the harmonic and radial oscillators in order to obtain new systems ruled by second and third-order PHA. Are those the only possible systems that can be reduced to these algebras? We doubt it, and thus we would be interested in identifying new systems ruled by these PHA. Then through these reduction theorems we derived a method to obtain solutions to PIV and PV equations in terms of the confluent hypergeometric function in certain subspace of the parameter space of the Painlev\'e equations. In this topic, we think it would be important to study further the general structure of these solutions, e.g., their node distribution and asymptotic behaviour. Furthermore, we have only scratched the surface of the explicit solutions, and we think it will be useful to explore deeper the method.

Finally, we classified these solutions into different solution hierarchies. In this subject, we think that a more detailed classification lie still deeper into their structure.


\section*{Acknowledgement}
The authors acknowledge the financial support of Conacyt (Mexico) project 152574. DB also acknowledges the Conacyt Ph.D. scholarship 219665.

\bibliographystyle{plain}

\end{document}